\documentclass[fleqn,usenatbib]{mnras}  
\usepackage{graphicx}	
\usepackage{amsmath}	
\usepackage{amssymb}	
\usepackage{makecell}

\newcommand{\norme}[1]{\left\Vert #1 \right\Vert}
\newcommand{\para}[1]{\left(#1\right)}
\newcommand{\cro}[1]{\left[#1\right]}
\newcommand{\aver}[1]{\left\langle #1 \right\rangle}

\newcommand{\rz}{r_0}
\newcommand{\lz}{L_0}
\newcommand{\cnh}{C_n^2(h)}
\newcommand{\otf}[1]{\tilde{h}_{#1}}

\newcommand{\psd}[1]{\mathcal{W}_{#1}}
\newcommand{\cov}[1]{\mathcal{B}_{#1}}
\newcommand{\rhol}{\boldsymbol{\rho}/\lambda}
\newcommand{\rhovec}{\boldsymbol{\rho}}
\newcommand{\gao}{g_\text{ao}}
\newcommand{\gtt}{g_\text{tt}}
\newcommand{\gal}{g_\text{al}}
\newcommand{\sinc}[1]{\text{sinc}\left(#1\right)}
\newcommand{\covRes}{\mathcal{B}_\varepsilon}
\newcommand{\covPerp}{\mathcal{B}_\perp}
\newcommand{\covPara}{\mathcal{B}_\parallel}
\newcommand{\covAO}{\mathcal{B}_\text{AO}}
\newcommand{\covTT}{\mathcal{B}_\text{TT}}
\newcommand{\covAL}{\mathcal{B}_\text{AL}}
\newcommand{\covAniso}{\mathcal{B}_\Delta}
\newcommand{\cobm}[1]{\textcolor{black}{#1}}
\newcommand{\cred}[1]{\textcolor{black}{#1}}

\title[PSF reconstruction for SPHERE/ZIMPOL]{Pushing Point spread function reconstruction to the next level. Application to SPHERE/ZIMPOL.}

\author[O. Beltramo-Martin et al.]{
	O. Beltramo-Martin$^{1,2}$\thanks{E-mail: olivier.beltramo-martin@lam.fr},
	A. Marasco$^{3,4}$,		
	T. Fusco$^{1,2}$	,
	D. Massari$^{5,3,4}$,	
	J. Milli$^{7,8}$	
	\newauthor
	G. Fiorentino$^{6}$,		
	B. Neichel$^{2}$	,
	\\
	$^{1}$ONERA, The French Aerospace Laboratory  BP. 72, F-92322 Chatillon Cedex, France\\
	$^{2}$Aix Marseille Univ., CNRS, CNES LAM, 38 rue F. Joliot-Curie, 13388 Marseille, France\\			
	$^{3}$ INAF - Osservatorio di Astrofisica e Scienza dello Spazio di Bologna, Via Gobetti 93/3, I-40129 Bologna, Italy\\	
	$^{4}$ Kapteyn Astronomical Institute, University of Groningen, NL-9747 AD Groningen, the Netherlands\\	
	$^{5}$ Dipartimento di Fisica e Astronomia, Universit\`{a} degli Studi di Bologna, Via Gobetti 93/2, I-40129 Bologna, Italy\\	
	$^{6}$ INAF-Osservatorio Astronomico di Roma, via Frascati 33, 0040 Monte Porzio Catone, Italy\\
	$^{7}$ European Southern Observatory (ESO), Alonso de C\'ordova 3107, Vitacura, Casilla 19001, Santiago, Chile  \\	
   	$^{8}$ Univ. Grenoble Alpes, CNRS, IPAG, F-38000 Grenoble, France
}

\date{Accepted XXX. Received YYY; in original form ZZZ}
\pubyear{2019}

\begin{document} 
	\label{firstpage}
	\pagerange{\pageref{firstpage}--\pageref{lastpage}}
	\maketitle
	\begin{abstract}
		\cobm{Point spread function (PSF) reconstruction (PSF-R) is a well established technique to determine reliably and accurately  the PSF from Adaptive Optics (AO) control loop data. We have successfully applied this technique to improve the precision on photometry and astrometry to observation of NGC6121 obtained with SPHERE/ZIMPOL as it will be presented in a forthcoming letter. Firstly, we present the methodology we followed to reconstruct the PSF combining pupil-plane and focal-plane measurements using using our PSF-R method PRIME \citep{BeltramoMartin2019_PRIME}, with upgrade of both the model and best-fitting steps compared to previous papers. Secondly, we highlight that PRIME allows to maintain the PSF fitting residual below 0.2\% over 2 hours of observation and using only 30\,s of AO telemetry, which may have important consequences for telemetry storage for PSF-R purpose on future 30-40\,m class telescopes.  Finally, we deploy PRIME  in a more realistic regime using faint stars so as to identify the precision needed on the initial guess parameters to ensure the convergence towards the optimal solution.} 
	\end{abstract}	
	\begin{keywords}
		instrumentation: Adaptive optics -- atmospheric effects -- methods:data analysis --  methods: analytical
	\end{keywords}
		
\section{Introduction}
\label{S:Intro}
	
Point Spread Function (PSF) analysis of seeing and diffraction limited images in crowded stellar fields became a consolidate technique since the late eighties. There are several commonly used astronomical packages that successfully apply this technique, such as STARFINDER \citep{ Diolaiti2000}, SEXTRACTOR \citep{Bertin1996}, or DAOPHOT \citep{Stetson1987}. These are essentially based on the extraction and the modeling of the PSF using isolated stars across the scientific Field of View (FoV). This task can be particularly challenging in two opposite cases, in very dense stellar fields, like the core of globular clusters, where a solid PSF modeling is compromised by confusion, as well as in sparsely dense fields surrounding isolated galaxies, where no or few point-like sources are available for PSF modeling. These two situations become even more demanding when observing with Adaptive Optics (AO) instruments since the size of the imaged FoV ranges typically from tens of arcsec to few arcmin, thus preventing us very often to have isolated stars at disposal for a robust PSF analysis. Furthermore, when dealing with AO systems, the complexity of the PSF increases when compared with seeing limited images and, last but not least, the real time turbulence correction make the PSF changing across the FoV, the time and the spectrum. The combination of source crowding and PSF inhomogeneity strongly affects the estimate of source parameters like magnitude and position \citep{Massari2016_a,Massari2016_b,Shodel2010,Yelda2010}. Photometric and astrometric accuracy can be enhanced using innovative approaches that account for AO PSF spatial variations models \citep{Ciurlo2018,Witzel2016}.

In the new era of Giant Segmented Mirror Telescopes (GSMTs), mainly fed by AO modules, our ability in performing good PSF fitting will be crucial. In fact, we aim to improve the photometric and astrometric accuracy and precision by at least a factor three, while the complexity of the PSF structure will increase. In this context, an alternative PSF determination approach, the so-called PSF reconstruction (PSF-R), is investigated since 20 years. This method uses information from the AO control loop data in order to build a theoretical PSF model as reminded in Sect. \ref{S:Model}. As analyzed in \citet{Ascenso2015}, PSF-R is truly promising to tackle actual limitations of standard standalone image processing pipeline. However, PSF-R has never reached the point to be fully integrated into a dedicated software for image analysis and operable for astronomers, calling for pushing PSF-R to a more suitable mode for later implementation into a pipeline.

\cobm{Moreover, one of the major issues PSF-R developers experienced is the calibration of scalar system parameters, such as seeing and WFS optical gains for instance. Although it remains feasible to identify these parameters from the telemetry, estimates are usually provided with 10\% accuracy \citep{Jolissaint2018}, which limits the PSF-R accuracy to closely at the same percentage as we illustrate in Sect. \ref{S:NGC6121}. Another way to say it is that handling the sole AO telemetry alone is not sufficient to reach 1\% level accuracy on the PSF. However, a very accurate PSF estimates at 1\% level may be needed, especially for stellar populations analysis  In this context, we have introduced PRIME \citep{BeltramoMartin2019_PRIME} as a novel approach that combines pupil-plane (AO telemetry) and focal-plane (imager frames) data to overcome the problem of system parameters identification and therefore optimize the PSF-R.}

\cobm{We have successfully deployed PRIME on NGC6121 images obtained in 2018 with the Zurich IMaging POLarimeter (ZIMPOL) instrument \citep{Schmid2018} mounted at the focal plane of the Spectro Polarimetric High-contrast Exoplanet REsearch (SPHERE) instrument \citep{Beuzit2019} at the Very Large Telescope (VLT). Results showed a factor 10 improvement on photometric precision and will be presented in an upcoming paper \citep{Massari2020}. Firstly, we are willing in this paper to present the exact methodology we followed to reconstruct the PSF using an upgraded version of PRIME presented  in Sect.\ref{SS:PRIME}. Secondly, another major drawbacks of PSF-R for future GSMTs lies in the large amount of telemetry data we must handle and archive every observation. With PRIME, we show in Sect. \ref{SS:onaxisPRIME} that we can mix 30\,s of telemetry data with images acquired 2\,h later and still achieve an accurate PSF determination below 0.2\% of mean residual, which offers the possibility to not store the full AO telemetry during observation. Finally, in Sect. \ref{S:offaxisPRIME}, we present an utilization of PRIME in a more realistic scenario using faint stars (mV=15-16 mag) and compared the reconstructed PSF with the on-axis bright source (mV=10.6 mag) of the NGC6121 field. In order to mitigate noise propagation into the criterion solving, we compare different methods to constrain the solution using either hard-bounds or Gaussian statistics assumption on parameters and conclude about the best         strategy. Discussions and conclusions are given in Sect. \ref{S:conclusion}.}

\section{Reconstructing the PSF}
\label{S:Model}

PSF-R inherits from image formation theory proposed by \citet{Roddier1981} to connect the focal-plane image to the incoming wavefront distortions within the pupil plane. In the case we neglect scintillation effect and assume that the phase of the electric field in the pupil plane is spatially stationary, the \cobm{long-exposure (e.g. the acquisition time is much larger than the turbulence coherence time)} Optical Transfer Function (OTF), defined as the PSF Fourier transform, is decomposed as the following multiplication:
\begin{equation}
\otf{}(\rhol) = \otf{\text{T}}(\rhol).\otf{\varepsilon}(\rhol).
\label{E:otf}
\end{equation}
where $\rhol$ is the angular frequency with $\lambda$ the observation wavelength, $\otf{}$ the total OTF, $\otf{T}$ the
telescope + instrument OTF that is derived from the pupil function autocorrelation and $\otf{\varepsilon}$ the residual atmospheric OTF. PSF-R aims at estimating $\otf{\varepsilon}$ from
\begin{equation}
\otf{\varepsilon}(\rhol) = \exp{\para{\covRes(\rhovec) - \covRes(0)}},
\label{E:otffit}
\end{equation}
where $\covRes$ is the residual phase covariance function that can be theoretically captured from the AO control loop data, i.e. the Wavefront Sensor (WFS) measurements and commands applied to the Deformable Mirror (DM). $\cov{\varepsilon}$ is split into a sum of covariance error terms, assumed to be independent each other as follows
\begin{equation}
\begin{aligned}
\covRes(\rhovec) &= \covPerp(\rhovec) + \covPara(\rhovec) + \covAniso(\rhovec),
\end{aligned}
\label{E:cov}
\end{equation}
$\covPerp$ refers to uncompensated spatial frequencies while $\covPara$ include AO residual only and $\covAniso$ corresponds to the anisoplanatism effect. The calculation of this latter can be found in multiple references \citep{BeltramoMartin2018_aniso,Britton2006,Flicker2003,Fusco2000,Whiteley1998}. In the case of SPHERE/ZIMPOL, we must only account for angular anisoplanatism that is produced by the spatial decorrelation of the incoming wavefront. In V-band, the typical isoplanatic angle at Paranal has been measured to $\theta_0 = 2$\,arcsec \citep{Osborn2018,Masciadri2014}, which is the separation from the guide star beyond which the PSF becomes significantly elongated by the anisoplanatism effect.

\subsection{Reconstruction of uncompensated modes}
$B_\perp(\rhovec)$ is calculated from the perpendicular Power Spectrum Density (PSD) as follows
\begin{equation}
\covPerp(\rhovec) = \mathcal{F}^{-1}\cro{\psd{\perp}(\boldsymbol{k}/\lambda)},
\end{equation}
where $\mathcal{F}^{-1}$ is the Fourier operator and $\psd{\perp}$ is the Von-K\' arm\' an atmospheric PSD filtered by the corrected frequencies
\begin{equation}
\psd{\perp}(\boldsymbol{k}) = \left\lbrace 
\begin{aligned}
& 0.0229\rz^{-5/3}(k^2 + 1/\lz^2)^{-11/6} \text{ if } k>k_\text{AO}\\
& 0 \text{ \hspace{3.8cm} otherwise}
\end{aligned}
\right.
\label{E:psdFit}
\end{equation}
with $\rz$ is the Fried's parameter (connected to the seeing), $k = |\boldsymbol{k}|$ and $k_\text{AO}$ is the AO cut-off frequency approximated by $k_\text{AO} \simeq n_\text{act}/2D$, where $n_\text{act}$ is the number of actuators per row and D the pupil diameter. This cut-off frequency is a function of the number of controlled modes, but we can still define an equivalent value of $n_\text{act}$ that produces the same fitting error, in the sense that this approximation is not critical for the following. The fitting PSD can be calculated once and scaled regarding the $\rz$ value. 

\subsection{Reconstruction of corrected modes}

$\covPara$ is computed directly from the AO telemetry \citep{Veran1997}. However, some care must be given to the aliasing that affects WFS measurements and the PSF morphology. In the case of large bandwidth system (with a measurements rate of 1380\,Hz, SPHERE complies with this hypothesis), \citet{Veran1997} shows that the phase produced by the DM is anticorrelated to the WFS aliasing perturbation created by high-order modes. Therefore, calculations show that to account for aliasing in the PSF-R process, one must add the aliasing closed-loop covariance to the covariance function derived from WFS slopes (that is contaminated by the aliasing). Moreover, tip-tilt modes are corrected  by using a Linear Quadratic Gaussian (LQG) algorithm \citep{Petit2014}, while higher order modes are controlled using a modal optimization-based integrator \citep{Petit2008}. This does not matter in the present case, expect for noise variance modeling for faint guide stars, however we will have to handle tip-tilt modes differently from other corrected ones in Sect. \ref{S:PRIME} and we introduce herein a split reconstruction for latter use. Eventually, we have
\begin{equation}
\covPara(\rhovec) = \covAO(\rhovec) + \covTT(\rhovec)+  \covAL(\rhovec) 
\end{equation}
where $\covAO$ is the tip-tilt excluded AO residual phase covariance, $\covTT$ is the residual jitter covariance, $\covAL$ the model of the aliasing covariance that affects the PSF.

\subsubsection{Reconstruction of tip-tilt excluded modes}
Methodology for calculating $\covAO$ from WFS measurements is highly spread in the literature \citep{BeltramoMartin2019_PRIME, Ragland2018_PSFR,Jolissaint2015,Gilles2012,Veran1997}; practically we used the Vii algorithm proposed by \citep{Gendron2006} where
\begin{equation}
\covAO(\rhovec) = \sum_i^{n_\text{act}} \Lambda(i,i) V_{ii}(\rhovec),
\end{equation}
\cobm{being $V_{ii}$ the Vii functions obtained from the eigen decomposition of the matrix of the High-order DM (HODM) influence functions (HODM influence matrix). The use of the Vii functions allows us to speed up significantly the covariance calculation. } $\Lambda$ is the diagonal matrix that contains the eigenvalues of the covariance matrix $\mathcal{C}_\text{AO}$. This latter is estimated empirically from the tip-tilt excluded WFS measurements as follows \citep{Vigan2019__ZELDA_AA}
\begin{equation}
\mathcal{C}_\text{AO} = \para{\dfrac{2\pi}{\lambda}}^2\mathcal{R}_\text{AO}\para{\aver{\boldsymbol{s}.\boldsymbol{s}^t} - \aver{\boldsymbol{\eta}.\boldsymbol{\eta}^t}}\mathcal{R}_\text{AO}^t,
\end{equation}
where $\mathcal{R}_\text{AO}$ is the matrix that reconstructs the point-wise wavefront from the WFS measurements $\boldsymbol{s}$ that are contaminated by an additive noise $\boldsymbol{\eta}$ and where $\aver{\boldsymbol{x}.\boldsymbol{x}^t}$ refers to the empirical covariance matrix of vector $\boldsymbol{x}$. The noise covariance can be estimated using analytical formula \citep{Rousset1987}, from the measurements temporal autocorrelation \citep{Gendron1995} or from their temporal PSD \citep{Jolissaint2015} to mention only these approaches. In the specific case of SPHERE/ZIMPOL, the reconstructor $\mathcal{R}_\text{AO}$ is derived from successive multiplications of calibrated matrices:
\begin{equation}
\mathcal{R}_\text{AO} = \para{\dfrac{632.10^{-9}}{2\pi}}\boldsymbol{\text{iF}}.\boldsymbol{P}_\text{M2V}.\boldsymbol{P}_\text{S2M}.\boldsymbol{P}_\text{AO},
\end{equation}
where
\begin{itemize}
\item[$\bullet$] $\boldsymbol{\text{iF}}$ is the calibrated HODM influence matrix that converts HODM voltages to a wavefront at $632$ nm. It has a dimension
of $240^2 \times n_\text{act}$, where 240 is the pupil resolution in pixel during the calibration, which is interpolated to $2\times n_\text{act}+1$ for speeding up the reconstruction.
\item[$\bullet$] $\boldsymbol{P}_\text{M2V}$ is the mode-to-voltage matrix that transforms
$n_\text{mode}$ Karhunen-Lo\`eve (KL) modes to $n_\text{act}$ HODM voltages in stroke units.
\item[$\bullet$] $\boldsymbol{P}_\text{S2M}$ is the slope-to-mode matrix that projects $2\times n_\text{subap}$ slope measurements in pixel to $n_\text{mode}$ KL-mode.
\item[$\bullet$] $\boldsymbol{P}_\text{AO}$ is a $2n_\text{subap} \times 2n_\text{subap}$ matrix that filters the tip-tilt modes out from the slopes measurements.
\end{itemize}

\subsubsection{Reconstruction of tip-tilt modes}

The tip-tilt covariance function in Eq.\ref{E:cov} is derived as follows:
\begin{equation}
\begin{aligned}
\covTT(\rhovec) = \dfrac{1}{D^2}\left(\right.&\mathcal{C}_\text{TT}(1,1).\rho_x^2 + \mathcal{C}_\text{TT}(2,2)\rho_y^2\\
&\left. + 2\times\mathcal{C}_\text{TT}(1,2)\rho_x\rho_y\right),
\end{aligned}
\label{E:BTT}
\end{equation}
where $\rho_x,\rho_y$ are the x/y projection of the separation vector $\rho$ from $-D/2$ to $D/2$ and $\mathcal{C}_\text{TT}$ the tip-tilt covariance matrix.
\begin{equation}
\mathcal{C}_\text{TT} = \para{\dfrac{2\pi}{\lambda}}^2\mathcal{R}_\text{TT}\para{\aver{\boldsymbol{s}.\boldsymbol{s}^t} - \aver{\boldsymbol{\eta}.\boldsymbol{\eta}^t}}\mathcal{R}_\text{TT}^t,
\end{equation}
where $\mathcal{R}_\text{TT}$ permits to reconstruct the tip-tilt wavefront over the pupil in meter using
\begin{equation}
\mathcal{R}_\text{TT} = \dfrac{2.6\times \pi\times D}{180\times 3600}.(\boldsymbol{P}_\text{TT}.\boldsymbol{D}_\text{V2S})^\dagger.\boldsymbol{P}_\text{TT},
\end{equation}
with $(x)^\dagger$ the generalized invert matrix and
\begin{itemize}
\item[$\bullet$] $\boldsymbol{P}_\text{TT}$ is a $2 \times 2n_\text{subap}$ matrix that projects the slopes measurements onto tip-tilt modes.
\item[$\bullet$] $\boldsymbol{D}_\text{V2S}$ is the $2n_\text{subap} \times 2$ calibrated tip-tilt interaction matrix that converts the two tip-tilt DM commands into WFS slopes.
\end{itemize}

\subsubsection{Reconstruction of the aliasing}

The aliasing covariance function is derived from a model of the aliasing PSD $\psd{\text{AL}}$ \citep{Correia2014,Jolissaint2010,Flicker2008} that accounts for the spatial filtering of the Shack-Hartmann WFS and the temporal propagation of the aliased measurement through the AO loop. Eventually we have
\begin{equation}
\covAL(\rho) = \mathcal{F}^{-1}\cro{\psd{\text{AL}}(\boldsymbol{k})},
\end{equation}
with $\psd{\text{AL}}$ calculated for any considered atmospheric layer by, firstly aliasing the Von-K\'arm\'ann atmospheric PSD around multiple of $1/d$, with $d$ the sub-aperture size, secondly propagating the PSD through the Shack-Hartmann spatial filter \citep{Rigaut1998} and the temporal rejection function of the loop, which is spatialized by replacing temporal frequencies with $\boldsymbol{f}=\boldsymbol{k}.\boldsymbol{v}_l$ where $\boldsymbol{v}_l$ is the velocity vector (norm is the windspeed value, angle is the wind direction) of the l$^\text{th}$ layer. We obtain the following expression for the aliasing PSD in closed-loop operation
\begin{equation}
\begin{aligned}
\psd{\text{AL}}(\boldsymbol{k}) =&
\dfrac{0.0229\rz^{-5/3}}{4\sinc{\boldsymbol{k}d}}\times \sum_{\substack{p=-\zeta\\ p\neq 0}}^{\zeta} \sum_{\substack{q=-\zeta\\ q\neq 0}}^{\zeta}\sum_{l=1}^{n_\text{L}}\\
&f_l\dfrac{(\boldsymbol{k}^{-1}.\boldsymbol{k}_{pq})^2 \sinc{k_p d}\sinc{k_q d}}{\para{k^2 + 1/L_0(l)^2}^{11/6}}.\mathcal{H}_\text{cl}(l),
\end{aligned}
\label{E:psdalias}
\end{equation}
with
\begin{equation}
\begin{aligned}
&\mathcal{H}_\text{cl}(l) =\\
& \dfrac{g^2\sinc{k_pv_{lx}t_i}\sinc{k_qv_{ly}t_i}\exp(2i\pi(k_pv_{lx}  +k_qv_{ly})t_d)}{1-2(\iota-g)\cos\para{2\pi k_pv_{lx} t_i}\cos\para{2\pi k_qv_{ly} t_i} + (\iota-g)^2}
\end{aligned}
\label{E:Hcl}
\end{equation}
where
\begin{itemize}
\item[$\bullet$] $\boldsymbol{k}_{pq} = (k_p,k_q)$ with $k_p = k_x - p/d$ and $k_q =  k_y - q/d$ as the x/y frequency vectors shifted by respectively $p/d$ and $q/d$.
\item[$\bullet$] $\zeta$ is a unitless number defined by the highest frequency seen by the WFS normalized by the DM cut-off frequency $k_\text{AO}$. This number accounts for the spatial filtering in the WFS optical path that has been implemented to mitigate as much as possible the aliasing effect \citep{Poyneer2006}. 
\item[$\bullet$] $v_{lx} = v_l\cos(\omega_l)$ and $v_{ly} = v_l\sin(\omega_l)$ are the components of the velocity vector at height $h_l$, projected respectively on the x-axis and y-axis of the frequency plan, with $v_l,\omega_l$ the corresponding turbulence velocity and wind direction values at height $h_l$.
\item[$\bullet$] $L_0(l)$ is the outer scale vertical profile. Practically, we have chosen a flat profile and $L_0=25$, but methods exist to retrieve the integrated outer scale value from the telemetry \citep{Andrade2019}.
\item[$\bullet$] $t_i$ and $t_d$ are respectively the  WFS temporal sampling frequency and the loop delay, that is 2.3 frames (Cantalloube et al, in prep).
\item[$\bullet$] $g$ is the average of modal gain vector \citep{Petit2008}. To be more accurate, we should evaluate how the WFS aliasing propagate through each controlled KL mode and though the AO loop by taking into account the modal optimization. However such a description would drastically increase the complexity of the aliasing PSD computation for a small improvement eventually.
\item[$\bullet$] $\iota$ is the integrator leak factor set to 1 (no leak) in the rest of this paper.
\end{itemize}

\subsection{Application to SPHERE/ZIMPOL}
\label{S:NGC6121}

\subsubsection{Data}
\label{SS:data}

We have acquired observations of NGC6121 with the ZIMPOL V-filter (central wavelength 554\,nm, width 80.6\,nm) in the context of technical calibrations \footnote{ESO program ID of observations: 60.A-9801(S)} granted after the 2017 ESO calibration workshop \url{http://www.eso.org/sci/meetings/2017/calibration2017}.

\begin{figure}
\centering
\includegraphics[width = 8.5cm]{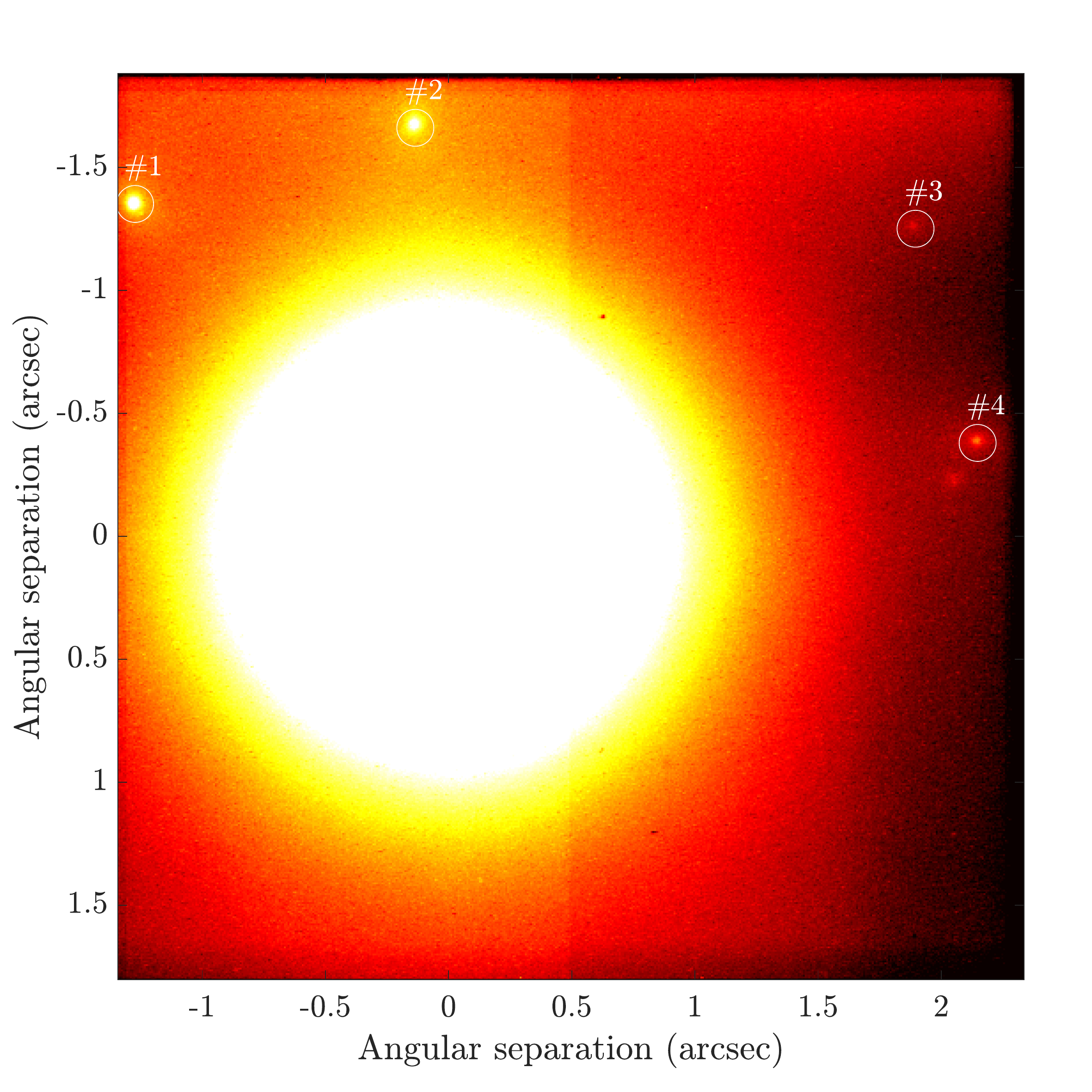}
\caption{\small Stack of the 26 frames of NGC6121 observations obtained with SPHERE/ZIMPOL on June 26th 2018. Coordinates correspond to the distance in arcsec from the bright AO guide star. The field was 3.5\,arcsec $\times$ 3.5\,arcsec large and four off-axis stars, encircled in white and numbered, were imaged.}
\label{F:NGC6121}
\end{figure}

With a pixel scale of 7.2\,mas/pixel, the detector was covering a 3.5\,arcsec $\times$ 3.5\,arcsec field of view as illustrated in Fig.\ref{F:NGC6121}. We summarize the acquisition time in Tab \ref{T:data} as the corresponding airmass value as well. The AO system ran at 300\, Hz, instead of the nominal 1380\, Hz owing to of the faintness (V=10.6\,mag) of the AO guide star and the grey dichroic used to share visible light between ZIMPOL and the WFS. The data were acquired in field stabilization with the Slow Polarimetry readout mode, which provides a readout noise 7 times smaller that the standard imaging mode, to enhance the S/N of the fainter off-axis sources. The data were reduced using the SPHERE Data and Reduction Handling pipeline (DRH) to extract the intensity image, subtract a  bias frame and correction for the flat-field. Dedicated Python routines were later used to recenter the individual frames, correct for bad-pixels and average the frames in a single reduced image as shown in Fig. \ref{F:NGC6121}.

The imaged field belongs to the central region of the aforementioned globular cluster, and includes five stars, the brightest of which was used to guide the AO system. The four others are more than 1.7\,arcsec away from the guide star. This field has been selected in order to quantify achievable limits on photometry and astrometry estimates of faint stars confused in the guide star halo and accurate HST measurements exist for comparison purpose.
 
\begin{table}
\centering
\caption{\small Summary of NGC6121 data successively acquired with ZIMPOL on the 26th June 2018 night, with a total of 26 images of 200s-exposure (NDIT=2, DIT=100s) each.}
\begin{tabular}{|c|c|c|c|}
\hline
Acquisition date & Airmass & Acquisition date & Airmass \\
\hline
\multicolumn{4}{|c|}{First slot of observations}  \\
\hline
\makecell{05:03:57 \\  + \textbf{AO data}} & 1.169 	& 05:26:06  & 1.238 \\
\hline
05:07:17 & 1.18 		& 05:29:26 &  1.243\\
\hline
05:11:27 & 1.191 	& 05:33:23  & 1.265  \\
\hline
05:14:47 & 1.201  	& 05:36:43 &  1.282 \\
\hline
05:18:49 & 1.213		& 05:42:11  & 1.300   \\
\hline
05:22:09 & 1.225		& 05:45:31 &  1.315   \\
\hline
\multicolumn{4}{|c|}{Second slot of observations} \\
\hline
\makecell{06:43:43 \\ + \textbf{AO data}} &  1.672  &07:08:57 & 1.934  \\
\hline
06:47:03 & 1.704	    & 07:12:55  & 1.977  \\
\hline
06:51:00 & 1.737   & 07:16:15  & 2.177    \\
\hline
06:54:20 & 1.771	   &07:20:19  &  2.077\\
\hline
06:58:18 & 1.809   &07:23:39   &  2.118 \\
\hline
07:01:38 & 1.849   &07:27:36  &  2.186\\
\hline
07:05:37 & 1.889   & 07:30:56 &  2.271\\  
\hline
\end{tabular} \label{T:data}
\end{table}

\begin{figure}
\centering
\includegraphics[width = 8.5cm]{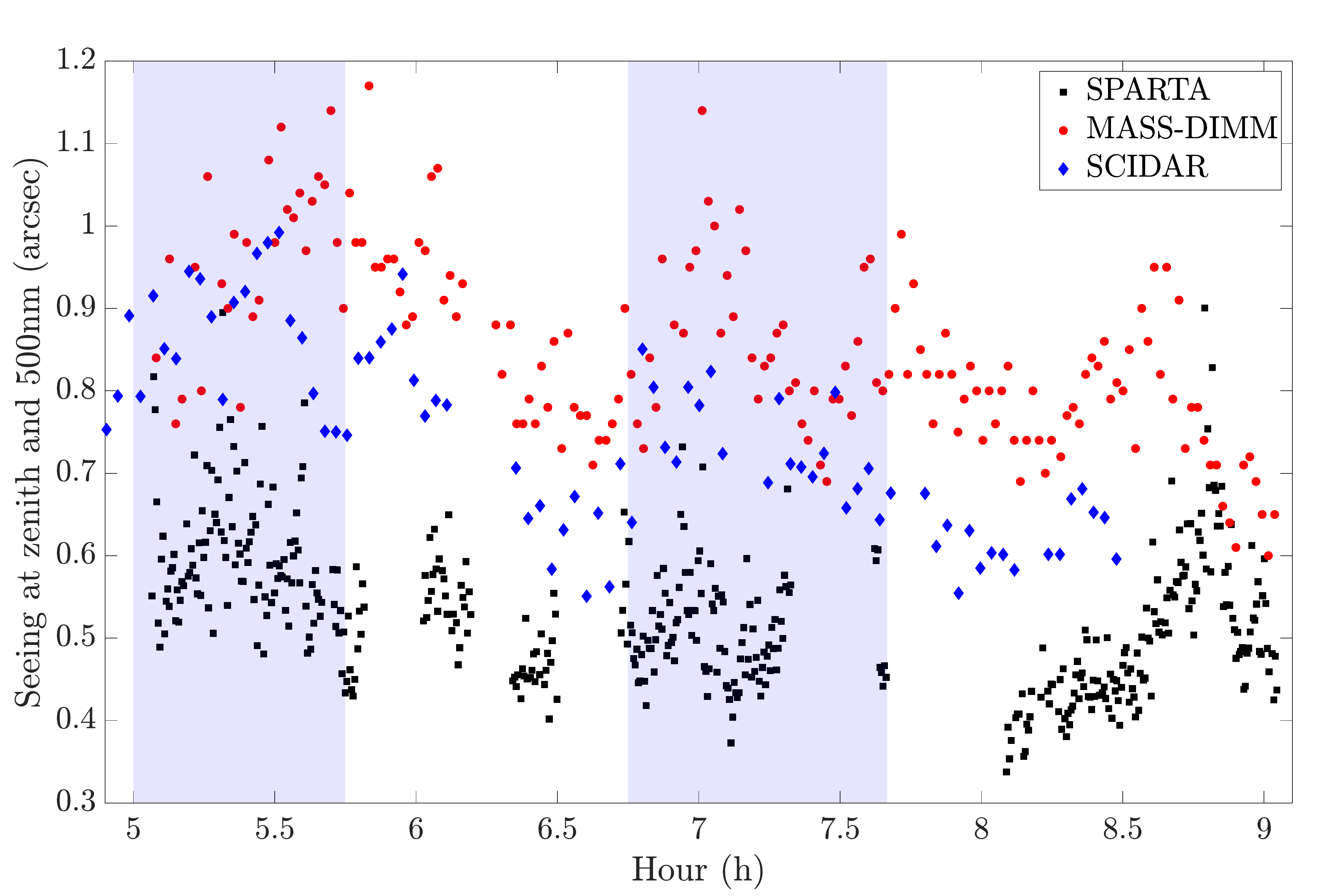}
\includegraphics[width = 8.5cm]{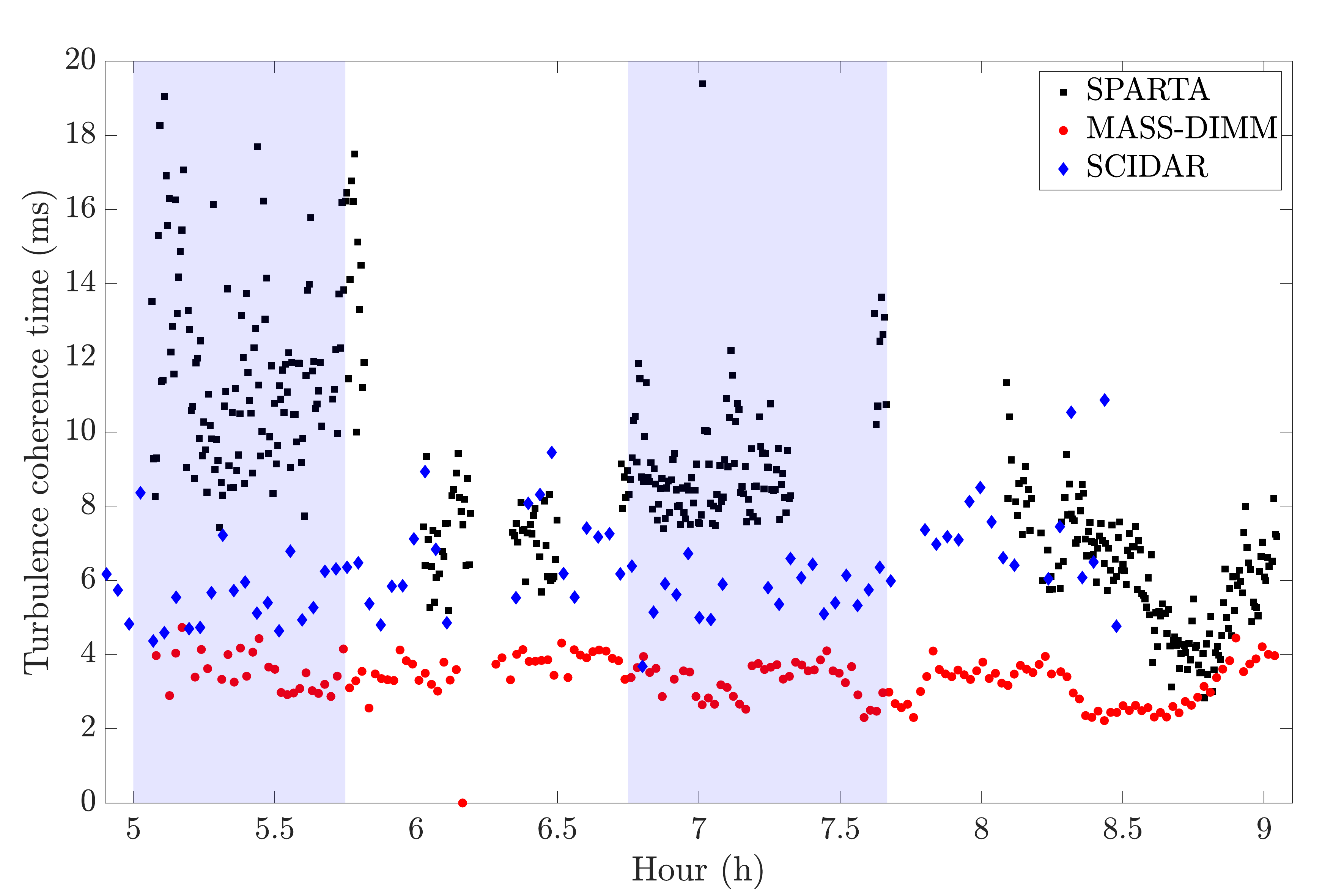}
\caption{\small \textbf{Top:} Seeing estimation at zenith and 500 nm \textbf{Bottom:} Turbulence coherence time estimation with respect to observing hour. Blue areas indicate the time window we had acquired ZIMPOL data.}
\label{F:seeing}
\end{figure}

On top of that, we had synchronous atmospheric parameters measurements delivered by, on the one hand, the MASS-DIMM at Paranal \citep{Butterley2018,Tokovinin2007} as well as the stereo-SCIDAR \citep{Osborn2018}, on the other hand, from the AO real time computer called SPARTA \citep{Fedrigo2006}. We present the temporal evolution during the observation in Fig.\ref{F:seeing}.  As already noticed by \citet{Milli2017}, there are large discrepancies between SPARTA and MASS-DIMM estimates, but both claim that observing conditions were quite stable
over the two observing slots. According to SPARTA, seeing and turbulence coherence time median values reached respectively 0.6\,arcsec and 11\,ms during the first part of the night, which evolved to 0.5\,arcsec and 9\, ms during the second slot.

Finally, we have also two 30\,s-long data sets of AO control loop data (WFS slopes, DM commands, calibrated matrices) obtained at 05h04m and 06h44m simultaneously with the beginning of ZIMPOL observations, as reported in Tab.\ref{T:data}. From the image, we have estimated a Strehl-ratio (SR) in V-band from the integration of the OTF over angular frequencies normalized to the diffraction limit OTF. We obtained of 1.4\,\%$\pm$ 0.63\,\% over the whole observation, which corresponds to 182\,nm $\pm$ 10, with a drop of 40\% in performance between slot 2 and slot 1.

\subsubsection{On-axis PSF-R results}
\label{SS:onaxisPSFR}

\begin{figure}
\centering
\includegraphics[width = 8.5cm]{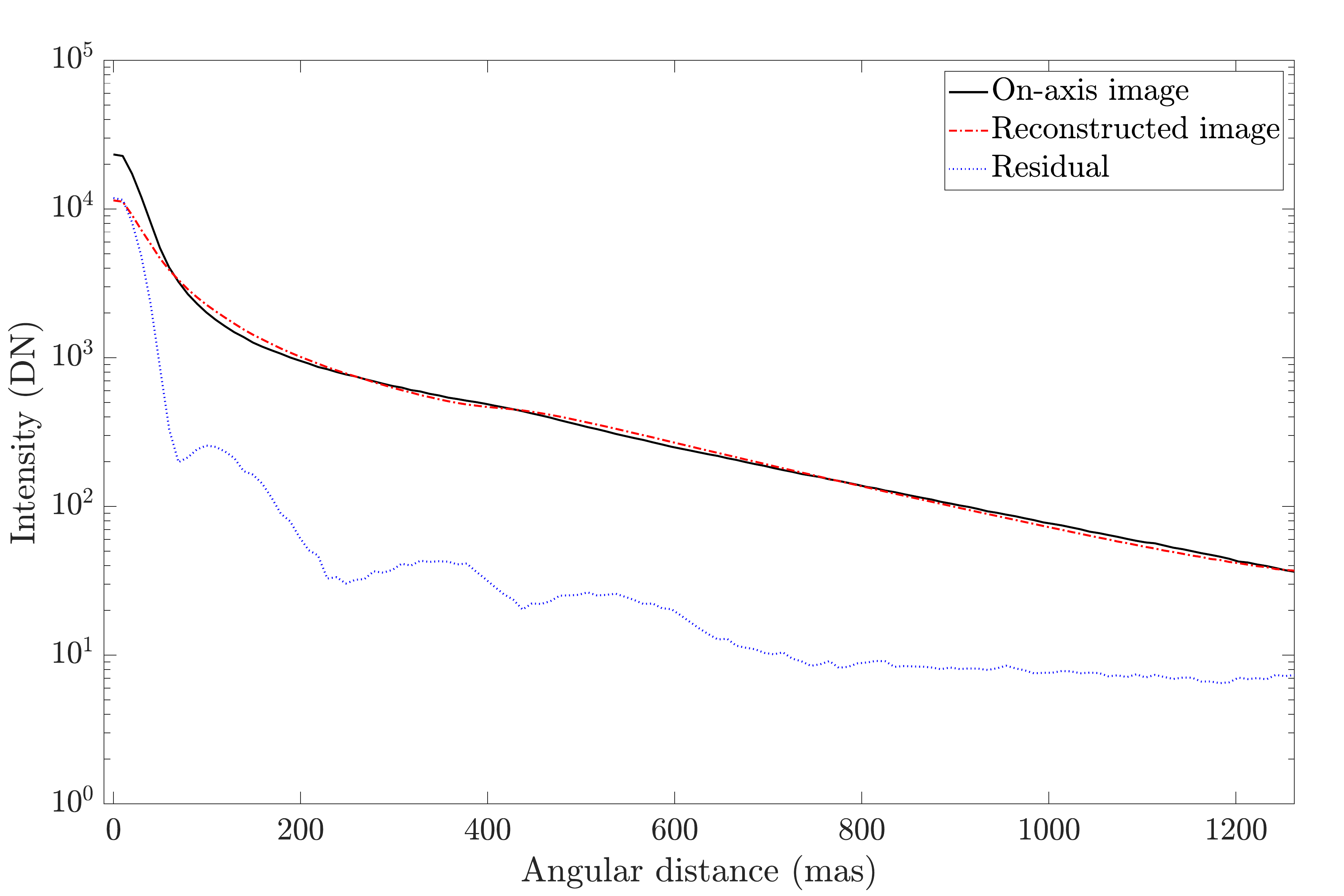}
\includegraphics[width = 8.5cm]{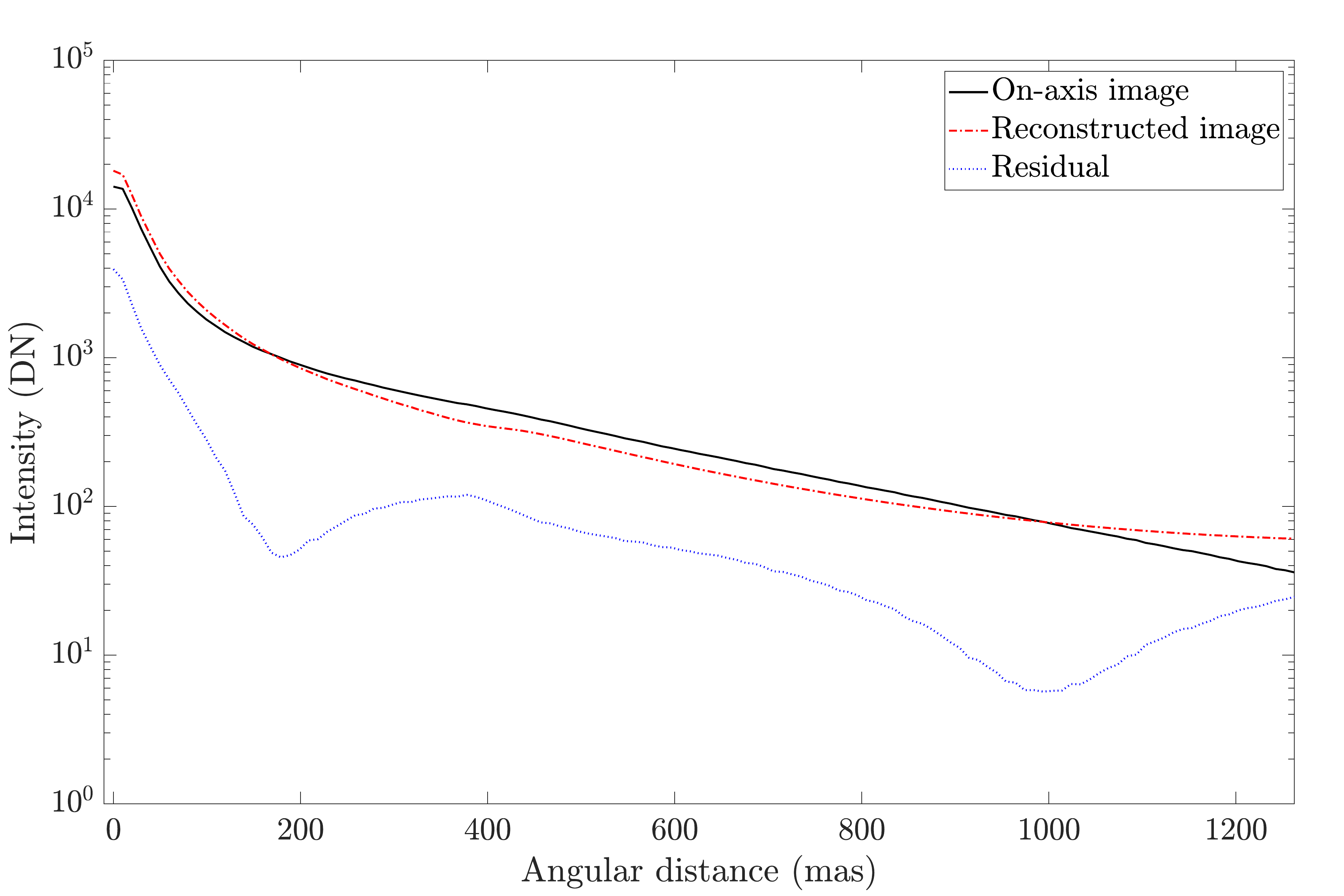}
\caption{\small Azimuthal average of sky and reconstructed PSFs obtained with 05:04 (\textbf{up}) and 06:43 (\textbf{down}) data. The PSF was reconstructed using the very same formalism presented in Sect. \ref{S:Model}. }
\label{F:forwardPSFR}
\end{figure}

We have followed the mathematical formalism presented in Sect. \ref{S:Model} to reconstruct the PSF from the two AO data sets acquired at the beginning of each observing time slot. The methodology to treat the AO control loop data was strictly the very same, and we present in Fig. \ref{F:forwardPSFR} reconstruction results compared to on-axis image. To obtain a proper comparison, we have adjusted the photometry and astrometry by using a weighted best-fitting to scale the PSF over the on-axis image. Results are somehow disappointing for several reasons. First of all, PSF wings are not systematically correctly retrieved, as we see with the second case for which $\rz$ is underestimated, ie. the atmospheric disturbances are expected to be stronger than they were actually. Moreover, the reconstruction of the PSF core does not behave similarly although the data processing is kept identical. We may have overestimation as well as underestimation of the PSF peak intensity. 

This situation occurs systematically when trying PSF-R: current algorithms usually fail achieving a stable, efficient and reproducible reconstruction across the multiple data sets they are tested on. There is a necessary  need to calibrate \textit{a posteriori} the algorithm over a sub-sample of data to approach the ultimate algorithm that would provide the same level of relative accuracy whatever the observing conditions. This calibration consists generally in applying some fudge factors to the telemetry \citep{Ragland2016,Martin2016JATIS,Clenet2008}, or/and to the noise variance. \textbf{The main conclusion of more than 20 years of efforts doing PSF-R is that PSF-R is not going to achieve proper PSF estimation by using the standard PSF-R framework.}

\subsubsection{Off-axis PSF-R results using SCIDAR data}
\label{SS:offaxisPSFR}

\begin{figure}
\centering
\includegraphics[width = 8.5cm]{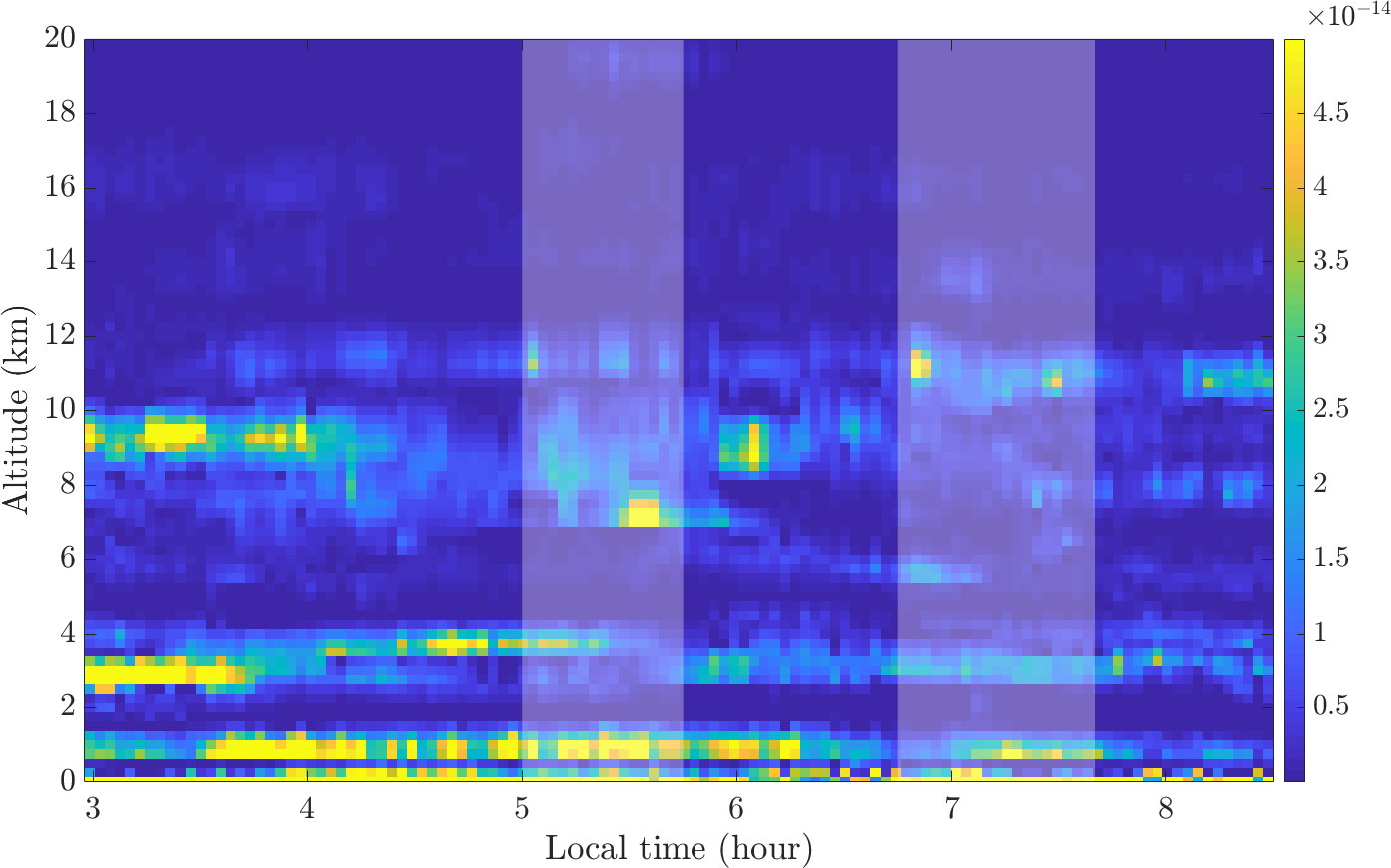}
\caption{\small $\cnh$ estimated by the SCIDAR at Paranal \citep{Osborn2018} across time. White bands corresponds to ZIMPOL observing time.}
\label{F:cn2_scidar}
\end{figure}

As illustrated in Fig. \ref{F:NGC6121}, the field also contains off-axis stars that are sufficiently distanced from the guide star to be potentially contaminated by the anisoplanatism effect. We have access to stereo-SCIDAR data at Paranal \citep{Osborn2018} acquired during the same time slot than ZIMPOL images. We report in Fig. \ref{F:cn2_scidar} the $\cnh$ evolution across time and the corresponding ZIMPOL observations time slots. This shows that the atmosphere was particularly concentrated into the first km, which does not produce a significant anisoplanatism. This latter was mostly generated by jet-streams between 8 and 12\,km, whose strength has slightly vanished across time.

 We report in Tab.~\ref{T:cn2} the $\rz$, average altitude and isoplanatic angle values obtained from the SCIDAR measurements at zenith and $\lambda = 500$\,nm. The $\rz$ value is calculated as follows \citep{Fried1965}
\begin{equation}
\label{E:rz}
\rz(h>H) = \para{0.423\para{\dfrac{2\pi}{\lambda}}^2\int_H^\infty \cnh dh}^{-3/5},
\end{equation}
where $H$ is the height above which the $\rz$ is measured from. The mean-weighted altitude $\bar{h}$ gives the layer height that would produce the same anisoplanatism regarding the $C_n^2$ distribution and is defined as \citep{Fried1982}
\begin{equation}
\label{E:hb}
\bar{h} = \para{\dfrac{\int_0^\infty h^{5/3}\cnh dh}{\int_0^\infty \cnh dh}}^{3/5}.
\end{equation}
Finally, the isoplanatic angle that defines the separation angle from the guide star above which the PSF evolves anisoplanatically is calculated from
\begin{equation}
\label{E:th0}
\theta_0 = 0.057\lambda^{6/5}\para{\int_0^\infty h^{5/3}\cnh dh}^{-3/5}.
\end{equation}

Values in Tab. \ref{T:cn2} show a mitigation of the anisoplanatism effect during the second observing slot, but because the large telescope zenith angle (60$\deg$), the S/N at off-axis stars directions has drastically diminished due to (i) atmospheric extinction that lessens the number of collected photons (ii) more anisoplanatism effect in the telescope line of sight that decreases the ensquared energy.
\begin{table}
\centering
\caption{\small Atmospheric characteristics given by the stereo-SCIDAR at zenith and 500\,nm.}
\begin{tabular}{|c|c|c|}
\hline
& First slot & Second slot \\
\hline
$\rz$ (cm) &  11.6 $\pm$ 1.1  & 13.8 $\pm$ 1.1\\
\hline
$\rz$(h$>$1\,km) & 15.3 $\pm$ 1.4 & 18.5 $\pm$ 1.9\\
\hline
$\bar{h}$ (km) & 6.4 $\pm$ 0.4 & 6.4$\pm$ 0.8 \\
\hline
$\theta_0$ (arcsec) & 1.2 $\pm$ 0.1 & 1.4$\pm$ 0.2 \\
\hline
$\theta_0$ (line of sight) & 1.1 $\pm$ 0.1 & 0.9$\pm$ 0.1 \\
\hline
\end{tabular}
\label{T:cn2}
\end{table}

Regarding $\theta_0$ and stars distance from on-axis, off-axis images should be slightly affected by the anisoplanatism effect. We have measured the PSF Full Width at Half Maximum (FWHM) of the five stars as reported in Tab. \ref{T:FWHM}, that indicates that the AO system did not reach the diffraction at 554\,nm, which was expected \citep{Fusco2014}. However, separation between the AO correction area and PSF wings is not very clear in Fig. \ref{F:PSFs}, advocating that there is an atmospheric or instrumental effect, which looks like to a residual jitter according to the elongated PSF pattern, that blurs the image. This is likely introduced by the low frame rate of 300\,Hz, which corresponds to the typical coherence time, that introduce a large servo-lag error. Consequently, off-axis PSFs are not significantly larger than the on-axis PSF, which suggests that the anisoplanatism is there, but the PSF morphology is dominated by the servo-lag error.

\begin{table}
\centering
\caption{\small PSFs FWHM for the five stars in the field derived from reconstructed PSF models.}
\begin{tabular}{|c|c|c|}
\hline 
& FWHM (mas) \\
\hline
Diffraction @ 554\,nm & 14  \\
\hline
On-axis PSF & 33.1 $\pm$ 2.0 \\
\hline
Off-axis PSF 1 & 37.4 $\pm$ 3.2 \\
\hline
Off-axis PSF 2 &  36.2 $\pm$ 2.8 \\
\hline
Off-axis PSF 3 &  38.9 $\pm$ 3.9\\
\hline
Off-axis PSF 4 &  38.6 $\pm$ 3.8 \\
\hline
\end{tabular}
\label{T:FWHM}
\end{table}

Our point is now to determine whether there is a need to account for anisoplanatism to calibrate the PSF model from off-axis stars. To do so, we have best-fitted off-axis images to retrieve photometry/astrometry using two different PSF models (i) the on-axis image (ii) the off-axis PSFs calculated by convolving the on-axis PSF with an anisoplanatism spatial filter calibrated from SCIDAR measurements. If there is a real anisoplanatism, the second model should give better results. Because S/N reasons, we have treated only the 12 first ZIMPOL frames that were acquired at a lower telescope zenith angle.

\begin{figure}
\centering
\includegraphics[width = 8.5cm]{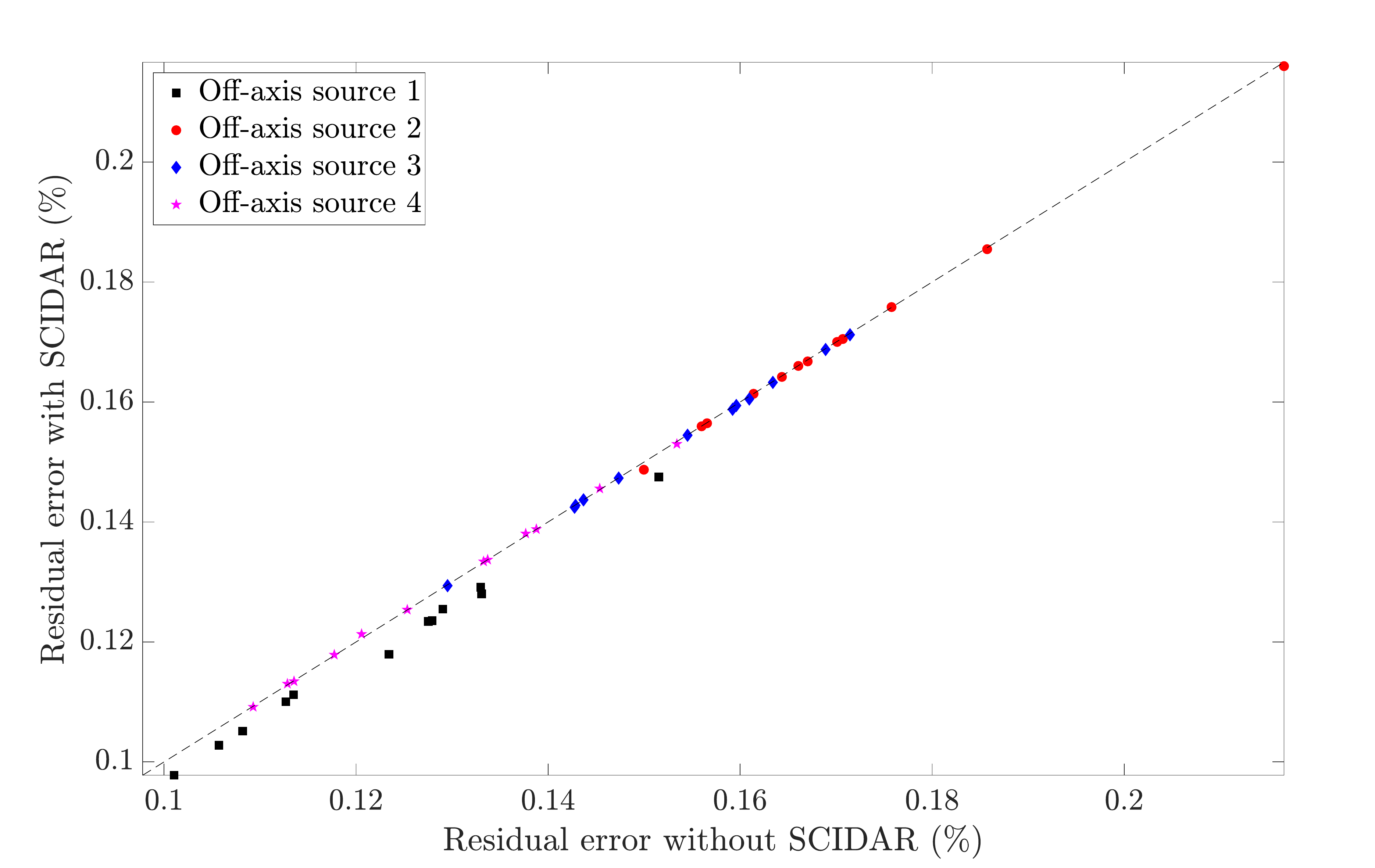}
\caption{\small Mean square error as calculated in Eq. \ref{E:res} obtained by best-fitting (Stellar parameters only) off-axis sources using a off-axis model (on-axis PRIME + SCIDAR) or the on-axis PRIME PSF. Results were obtained of the 12 first ZIMPOL frames.}
\label{F:eqmOff}
\end{figure}

In Fig \ref{F:eqmOff}, we compare the residual error obtained with Eq. \ref{E:res} computed with versus without use of SCIDAR data. We see that the off-axis PSF model barely improves the PSF model, but this improvement remains marginal enough to claim that the anisoplanatism does not contribute to PSF inhomogeneity, despite that the stars separation is larger than $\theta_0$. We show this in more detail for the off-axis star \#1 in Fig. \ref{F:eqmOff}: clearly, the residuals do not improve by including the effect of anisoplanatism. 
\cobm{This is likely due to a large residual jitter effect in the PSF that creates an elongated pattern that masks the anisoplanatism feature. In other words, accounting for the anisoplanatism produces only a very marginal improvement on the PSF modeling; therefore we chose to consider the PSF homogeneous across the field, which means that we will derive the PSF from the integrated $\rz$ value only.}



\begin{figure*}
\centering
\includegraphics[width = 17.5cm]{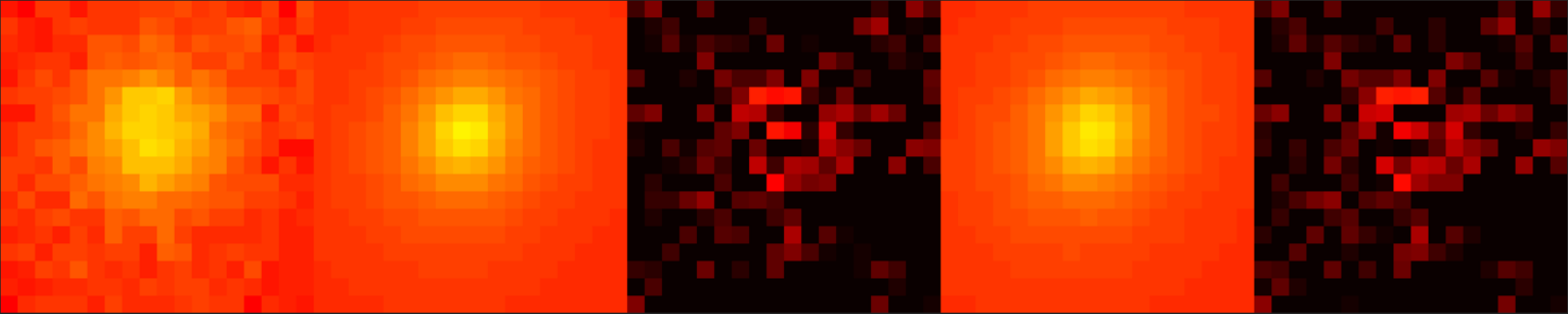}
\caption{\small 2D comparison of PSFs using a hyperbolic arcsinus scale. From left to right: off-axis star 1 image, on-axis PSF + SCIDAR, corresponding residual, PRIME without SCIDAR, corresponding residual. Reconstructed PSFs were obtained using the 05:04 AO data set.  }
\label{F:offPSFs}
\end{figure*}

\section{PRIME: a complementary tools to improve classical PSF-R scheme}
\label{S:PRIME}

\subsection{Introduction to PRIME}
\label{SS:PRIME}

New approaches must be envisioned to step into the next level of PSF-R. We have introduced the so-called method PRIME \citep{BeltramoMartin2019_PRIME} that yields a built-in parametric PSF model using AO measurements. As discussed, the reconstruction relies usually on fudge factors which are learned by comparing PSF-R to on-sky images and set up as constants afterwards. There were no tentative so far to figure out whether these fudge factors vary accordingly with observing conditions changes. 
This is what PRIME is going to achieve: it retrieves these factors by adjusting them to match available PSFs in the field. In other words, we combine pupil-plane (PSF-R framework) and focal-plane (best-fitting technique) to calibrate the PSF model across field and spectrum. PRIME permits to avoid collecting specific calibration data for PSF estimation purpose. For instance, there is no more need of offsetting the telescope or requesting technical time to test PSF-R.
The major drawbacks is the need of stars in the field, which does not comply with all science cases so far, but we will discuss in Sect. \ref{SS:products} what are the considered strategies to deliver a PSF regardless of the  science field configuration. Assuming that we have stars in the field, PRIME achieves the PSF calibration as follows
\begin{itemize}
\item[$\bullet$]  \textbf{Step 1:} Instantiate the PSF model by calculating the covariance matrices introduced in Sect. \ref{S:Model}.
\item[$\bullet$]  \textbf{Step 2:} Extract sub-fields of the image (user-defined) that contains one or several PSFs.
\item[$\bullet$] \textbf{Step 3:} Define PSF model parameters to be adjusted. In the following, we will estimate only four parameters using the following parametrization
\cobm{
\begin{equation}
\begin{aligned}
\covRes(\rhovec) &= \rz^{-5/3}\times(\covPerp(\rhovec,\rz=1)+ \gal\times\covAL(\rhovec,\rz=1))\\
& + \gao\times\covAO(\rhovec)+  \gtt\times\covTT(\rhovec) \\
\end{aligned}
\label{E:covPRIME}
\end{equation}}
where we assume to work in the isoplanatic regime, e.g. there is no need to include the anisoplanatism covariance function. On top of the $\rz$ value that will be estimated on the focal-plane image, we include three system parameters $\gao, \gtt, \gal$ that allow to play on the residual variance level. For a given AO data set, we calculate $\covAO$, $\covTT$ and $\covAL$, while $\covPerp$ is computed for once. \cobm{Contrary to the model presented in \cite{BeltramoMartin2019_PRIME}, we have introduced the gain $\gal$ to calibrate the aliasing model. The use of $\gal$ allows us to account for fluctuations in the wind speed profile, which enters the computation of the aliasing mode via Eq. \ref{E:psdalias} but can not readily be determined from telemetry data alone.}

\item[$\bullet$] \textbf{Step 4:} Minimizing a criterion by using a non-linear least-squares minimization algorithm (this is the current implementation, but others may be explored). We have used Matlab$^\circledR$  non-linear problem-fitting facilities based on a trust region reflective algorithm, as done by \citet{Fetick2019_Moffat_aa}, in order to minimize the following criterion
\begin{equation}
\label{E:crit}
\mathcal{J}(\boldsymbol{\mu},\gamma,\boldsymbol{\alpha}) = \sum_{i,j}^{n_\text{px}} w_{ij}\cro{\gamma \times \delta_{\boldsymbol{\alpha}}*h_{ij}({\boldsymbol{\mu}}) - d_{ij} + \nu}^2
\end{equation}
where $.*.$ is the convolution product and
\begin{itemize}
\item[$\bullet$] $h_{ij}$ and $d_{ij}$ are the $(i,j)$ pixel intensity values of respectively the numerical PSF model and sky observation. \cobm{The image is converted into e- units using an uniform detector gain of 1.5\,e-/ADU (slow Pol mode).},
\item[$\bullet$] $\boldsymbol{\mu} = [\rz,\gao,\gtt,\gal]$ is the set of parameters to be adjusted,
\item[$\bullet$] $\delta_{\boldsymbol{\alpha}}$ is the Dirac distribution shifted by the astrometric position $\boldsymbol{\alpha}$  and multiplied by the photometric factor $\gamma$,
\item[$\bullet$] $\nu$ is an additional  degree of freedom to account for a residual background,
\item[$\bullet$] $w_{ij}$ is the weighting coefficient for the $(i,j)$ pixel.
\end{itemize}

Identically to what has been done by \citet{Fetick2019_Moffat_aa} and \citet{Mugnier2004}, the weighting factor is set to
\begin{equation}
w_{ij} = \dfrac{1}{\text{max}\left\lbrace d_{ij},0\right\rbrace + \sigma_\text{RON}^2},
\end{equation}
where $\sigma^2_\text{RON}  =3$e- for the ZIMPOL detector (slow pol readout mode) \citep{Schmid2018}. \cobm{The detector noise is assumed to be Gaussian and independent from the data.} Also, Eq. \ref{E:crit} stresses that the model adjustment process does retrieve the PSF model parameters $(\rz,\gao,\gtt,\gal)$ and the stellar parameters $(\gamma, \boldsymbol{\alpha})$ simultaneously. This steps automatically output the astrometric/photometric measurements of calibration sources.
 
\item[$\bullet$] \textbf{Step 5:} Extrapolate the PSF to any desired field and spectrum position.
\end{itemize}

PRIME offers multiple applications, as image-assisted error breakdown  and photometry and astrometry measurements as discussed in \citet{BeltramoMartin2019_PRIME}. In this paper, we focus on the PSF estimation only. In order to quantify the quality of the PSF fitting, we will evaluate the relative mean square error given by
\begin{equation}
\epsilon_h = \dfrac{\sqrt{\sum_{i,j}^{n_\text{px}} \cro{\widehat{\gamma} \times \delta_{\boldsymbol{\widehat{\alpha}}}*{h}_{ij}(\widehat{\boldsymbol{\mu}}) - d_{ij} + \widehat{\nu}}^2}}{\sum_{i,j}^{n_\text{px}} d_{ij}}.
\label{E:res}
\end{equation}
where $\widehat{\gamma}$, $\boldsymbol{\widehat{\alpha}}$, $\widehat{\boldsymbol{\mu}}$ and $\widehat{\nu}$ are the estimated fitting parameters. The weighting process is particularly helpful to recover all spatial frequencies, without giving more importance to the AO corrected area where pixels are brighter. Especially, this ensures to mitigate biases in the $\rz$ estimation that depends significantly on the extended, low-intensity PSF wings. \cred{Note that this mean square error does not include the weight matrix as well as any regularization term we introduce in Sect. \ref{S:offaxisPRIME}. This metric allow to assess how far the reconstructed PSF, regardless the approach, is from the on-axis sky image and allow to analyze different techniques to each others regarding the residual error thay lead to.}

In the following, we will distinguish between \textit{forward PSF-R}, as the reconstruction process that relies on \textit{a priori} parameters ($\rz$ from telemetry, $\gao=\gtt=\gal=1$), and \textit{PRIME} as the PSF obtained after adjustment over the focal-plane image.

\subsection{Maximizing the utmost PSF-R performance}
\label{SS:onaxisPRIME}

We have tested PRIME by tuning PSF model parameters over the very bright on-axis star, i.e., the star guiding the AO system. The on-axis star is not necessarily present in the field, in particular for laser-assisted systems, in a way relying on this very bright source is not a nominal situation to deploy PRIME. However, the study of this on-axis star is optimal to test the utmost performance of PRIME, which will seek all the possible information on the PSF which are missing in the telemetry data directly on the image itself.
On top of that, the very accurate best-fit parameters found for the on-axis star can then be used as a reference and compared to those (less accurate) found for the faint off-axis sources, to assess how PRIME performs in less optimal situations as we do in Sect. \ref{S:offaxisPRIME}.
\cobm{Finally, in order to provide evidence that PRIME is a good solution for PSF-fitting problems post-AO, we have also compared herein results with a Moffat-fitting using the exact same model-fitting process. The Moffat function was defined over 7 parameters \citep{Fetick2019_Moffat_aa} to allow fitting of asymmetric PSF shape.}

\begin{figure}
\centering
\includegraphics[width = 8.5cm]{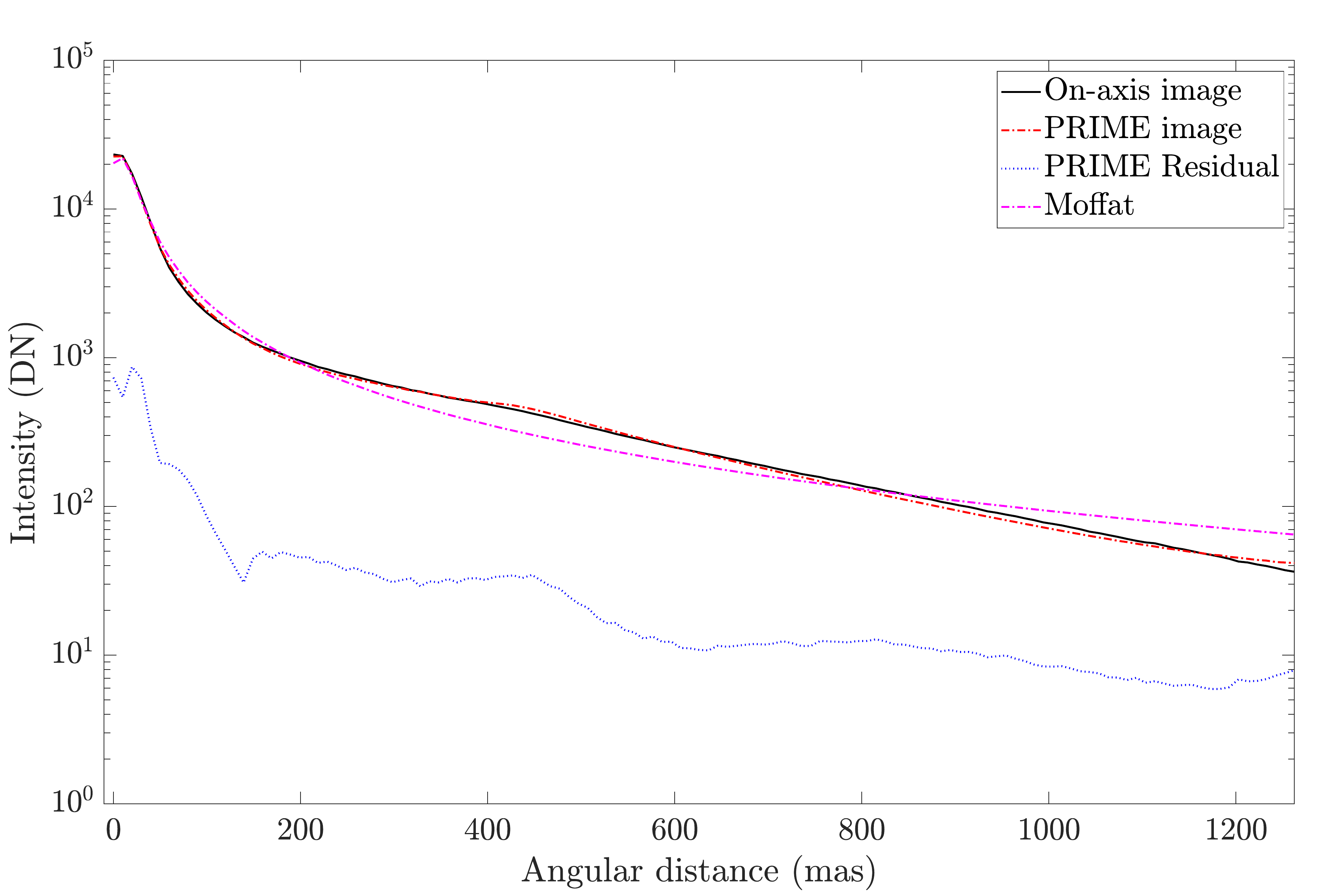}
\includegraphics[width = 8.5cm]{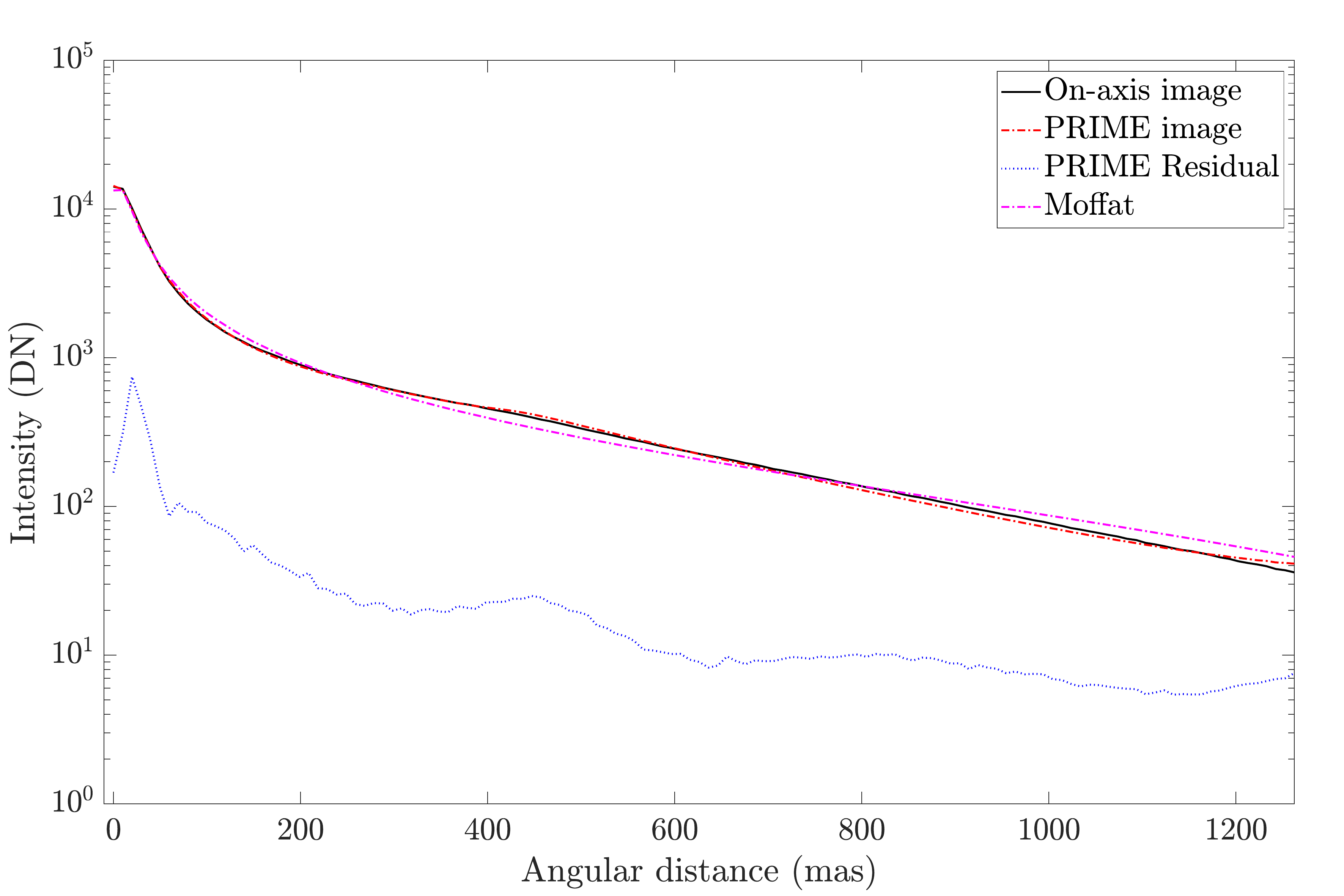}
\caption{\small  Azimuthal average of sky, PRIME and Moffat-fitted PSFs obtained with 05:04 (\textbf{up}) and 06:43 (\textbf{down}) data.}
\label{F:primePSFR}
\end{figure}

\begin{figure}
\centering
\includegraphics[width = 8.5cm]{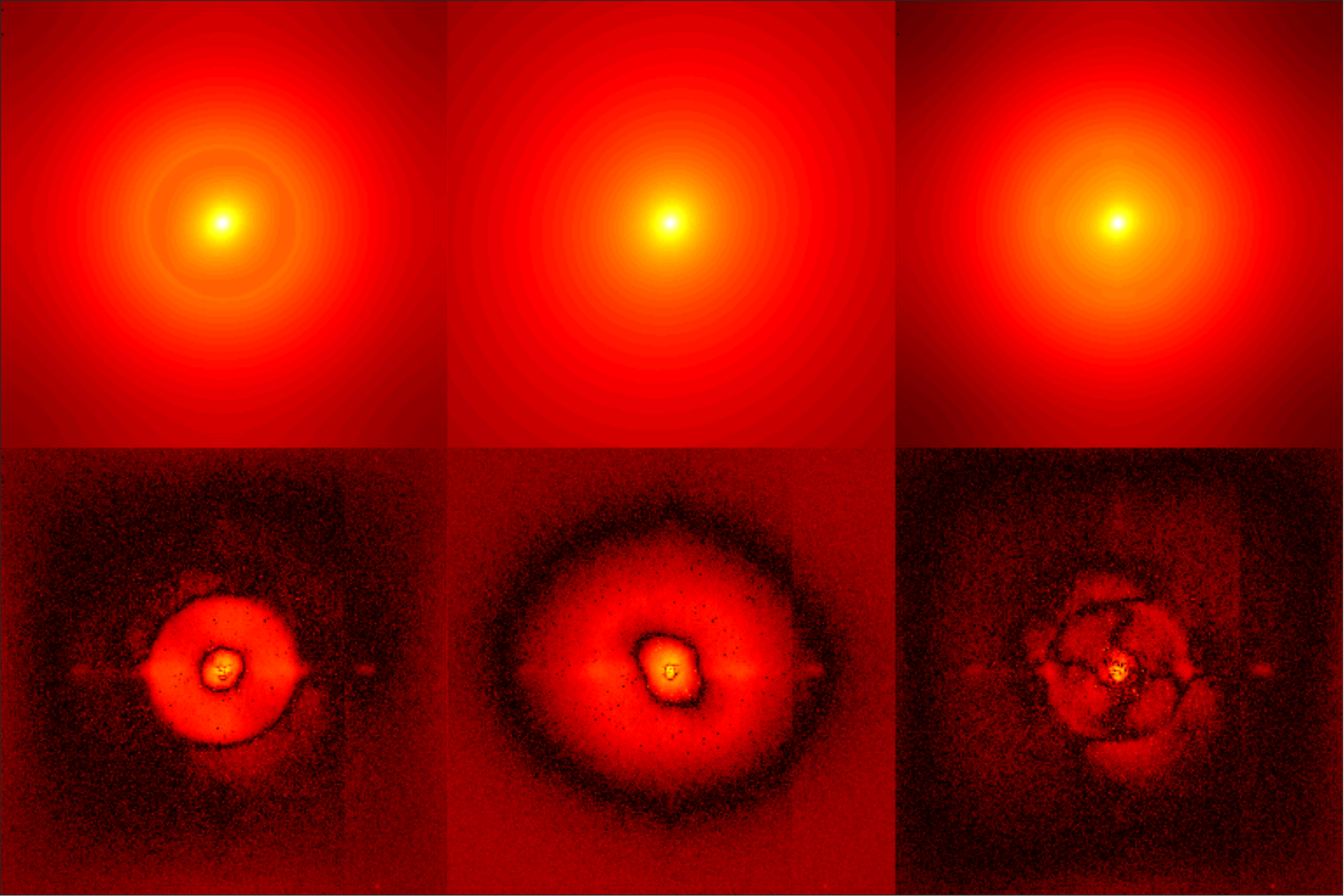}
\caption{\small \cobm{2D comparison of PSFs (05:03:57) using a hyperbolic arcsinus intensity scale. From left to right: forward PSF-R, Moffat-fitting, PRIME. Bottom images are residual obtained by subtracting the reconstruction to the sky observation.}}
\label{F:PSFs}
\end{figure}

We report in Fig. \ref{F:primePSFR} the comparison between the on-axis PSF profile with PRIME and Moffat-fitting as we did with forward PSF-R in Sect. \ref{SS:onaxisPSFR}. Moreover, we present a 2D comparison in Fig. \ref{F:PSFs} that compares forward PSF-R, PRIME and Moffat fitting and show a factor 2 at least improvement on the residual brought by PRIME, especially on the PSF core. The best-fitting achieves an excellent reproduction of the original image and in both cases, stressing that the calibration of additional factors on top of the AO telemetry treatment is a must. \cobm{ We also illustrate that the Moffat does not match the on-axis PSF as well as PRIME does, especially due to the PSF structure that contains an AO-corrected part and seeing-limited wings. We see an improvement of the Moffat-fitting on the second case for which the airmass was larger, though. In this situation, the line-of-sight seeing is worse than it was previously and the atmospheric residual is large enough to smooth the PSF and attenuate the transition sharpness between the AO corrected and non-corrected areas. However, the Strehl ratio was assessed at 1-2\% for these observations and at the imaging wavelength of 554\,nm. As a conclusion, even in a very poor correction regime, e.g. when the PSF structure is limited by atmospheric residual, it is already worth using PRIME instead of an analytical model that is designed to fit seeing-limited images.}

\subsection{Mitigating required data amount for PSF-R}
\label{SS:products}

So far we did not treat all the ZIMPOL frame we have at hand, only the one that have been synchronously acquired with AO telemetry. A question that appear when talking about PSF-R for next generation of AO-assisted instrument is related to the amount of data we need for performing PSF-R. Especially, there are some concerns about what it demands in terms of storage capability to record systematically all the AO telemetry and associated calibrated matrices.
With PRIME, we can address this question: do we necessarily need synchronous AO telemetry with science frame? To provide hints, we have utilized the two AO data sets to reconstruct two different PSF models using either forward PSF-R or PRIME. We end up with four PSF models. With forward PSF-R, the PSF was firstly reconstructed and then best-fitted (photometry/astrometry) to match the observation. with PRIME, both stellar and PSF parameters were jointly estimated through the model-fitting. 

\begin{figure}
\centering
\includegraphics[width = 8.5cm]{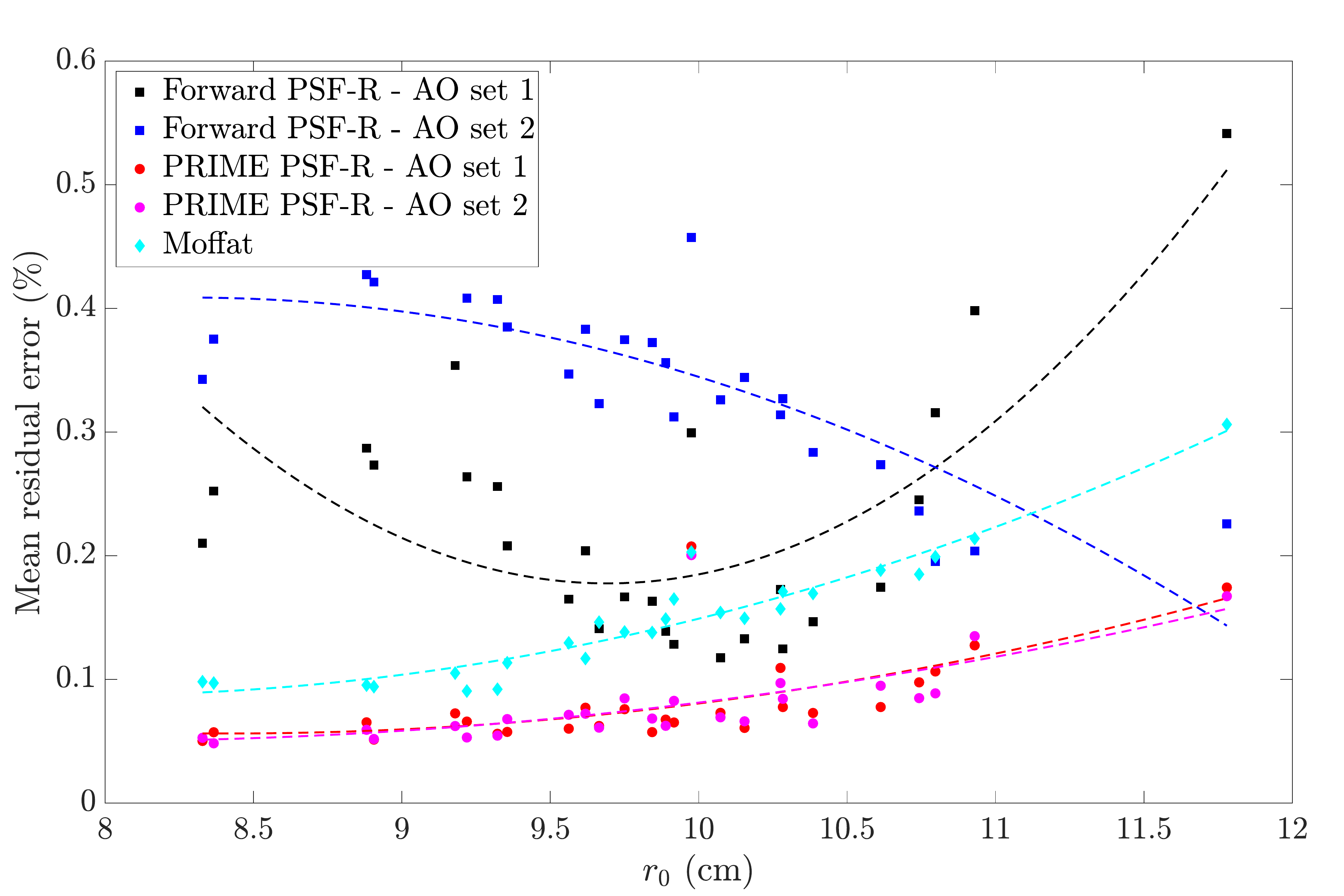}
\caption{\small \cobm{Mean square error as calculated in Eq. \ref{E:res} obtained with both forward PSF-R and PRIME models instantiated with either the 05:04 AO data set (set 1) or the 06:44 set (set 2), as well as the Moffat-fitting results. Dash lines gives quadratic trends with respect to the $\rz$ estimates obtained with PRIME}.   }
\label{F:eqmVr0}
\end{figure}

\cobm{We present in Fig. \ref{F:eqmVr0} the mean residual error as function of the estimated $\rz$ from PRIME and calculated from residual obtained using forward PSF-R, PRIME or a Moffat model. Several conclusions can be drawn. First of all, PRIME works much better than forward PSF and decreases the residual error by a factor 3 up to 10 over the 26 frames compared to forward PSF-R, and a factor 2 up to 4 compared to the Moffat model. Then, forward PSF-R efficiency degrades significantly when the AO telemetry is not synchronous with the imager frame, especially because variations of seeing conditions.  Finally, PRIME achieves a very stable reconstruction whose residual remains pretty much constant, even two hours after having recorded AO telemetry. In other words, temporal variability of atmospheric disturbances can be approximated as a scaling fluctuation of covariance functions introduced in Sect. \ref{S:Model}. Their structure remains quite similar across time, for these specific observations, though. }

\begin{figure}
\centering
\includegraphics[width = 8.5cm]{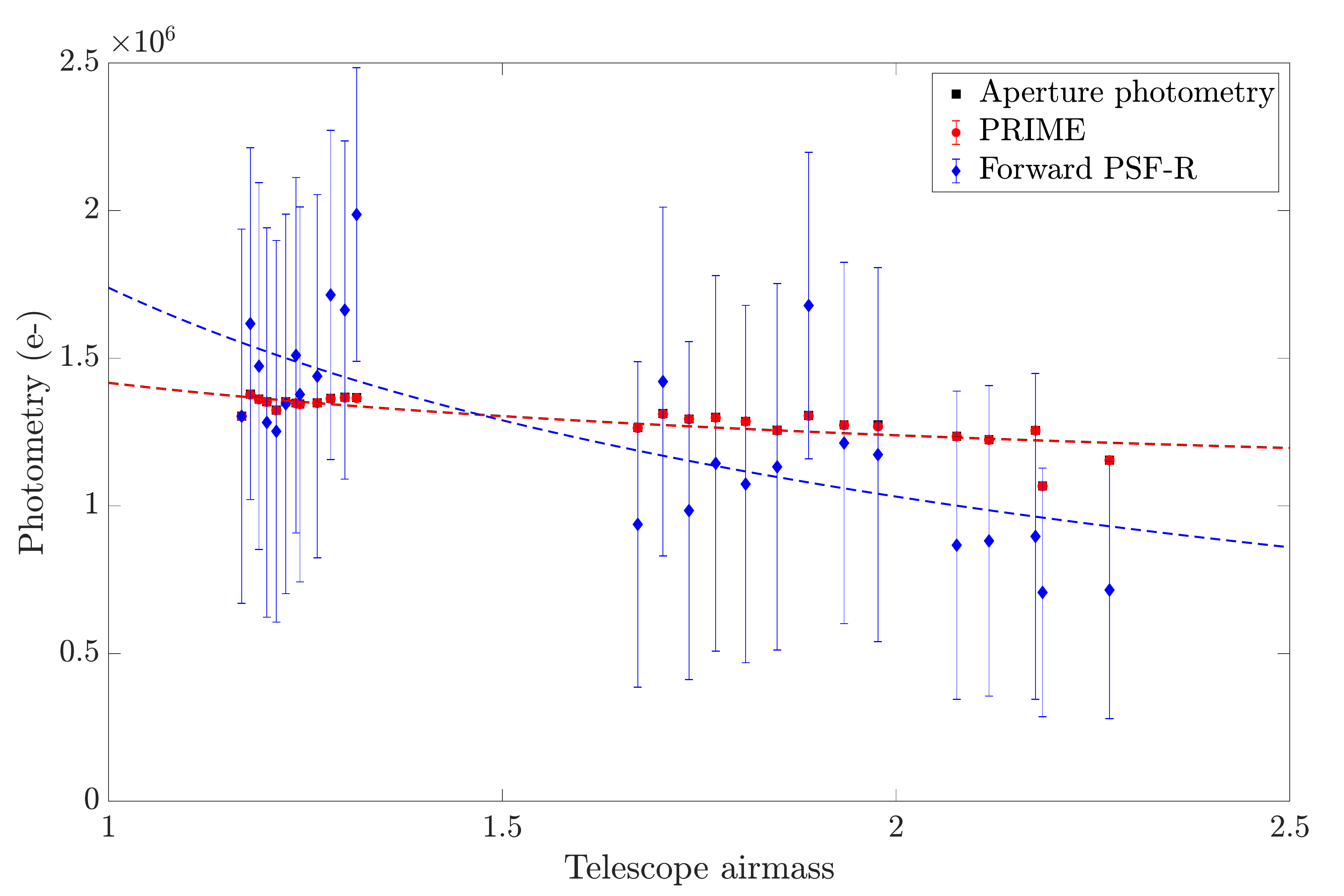}
\caption{\small Photometry measurements obtained on the 26 ZIMPOL frames as function of the telescope airmass. Dash lines refer to trends in $\text{airmass}^{-3/5}$. Error bars are obtained from the fitting minimization function.}
\label{F:photometryVairmass}
\end{figure}

On top of that, we report in Fig. \ref{F:photometryVairmass} photometry measurements using either forward PSF-R, PRIME and aperture photometry (sum of pixels). We clearly put into light that PRIME achieves accurate photometry measurements, while forward PSF-R is highly biased and imprecise. This is also indicated by the flux decreasing, due to atmospheric extinction and ensquared energy diminution, that does not decay with a similar slope. Precise assessment of photometry and astrometry capability of PRIME, with comparison to standard post-processing pipeline will be led at a later stage.

\begin{figure}
\centering
\includegraphics[width = 8.5cm]{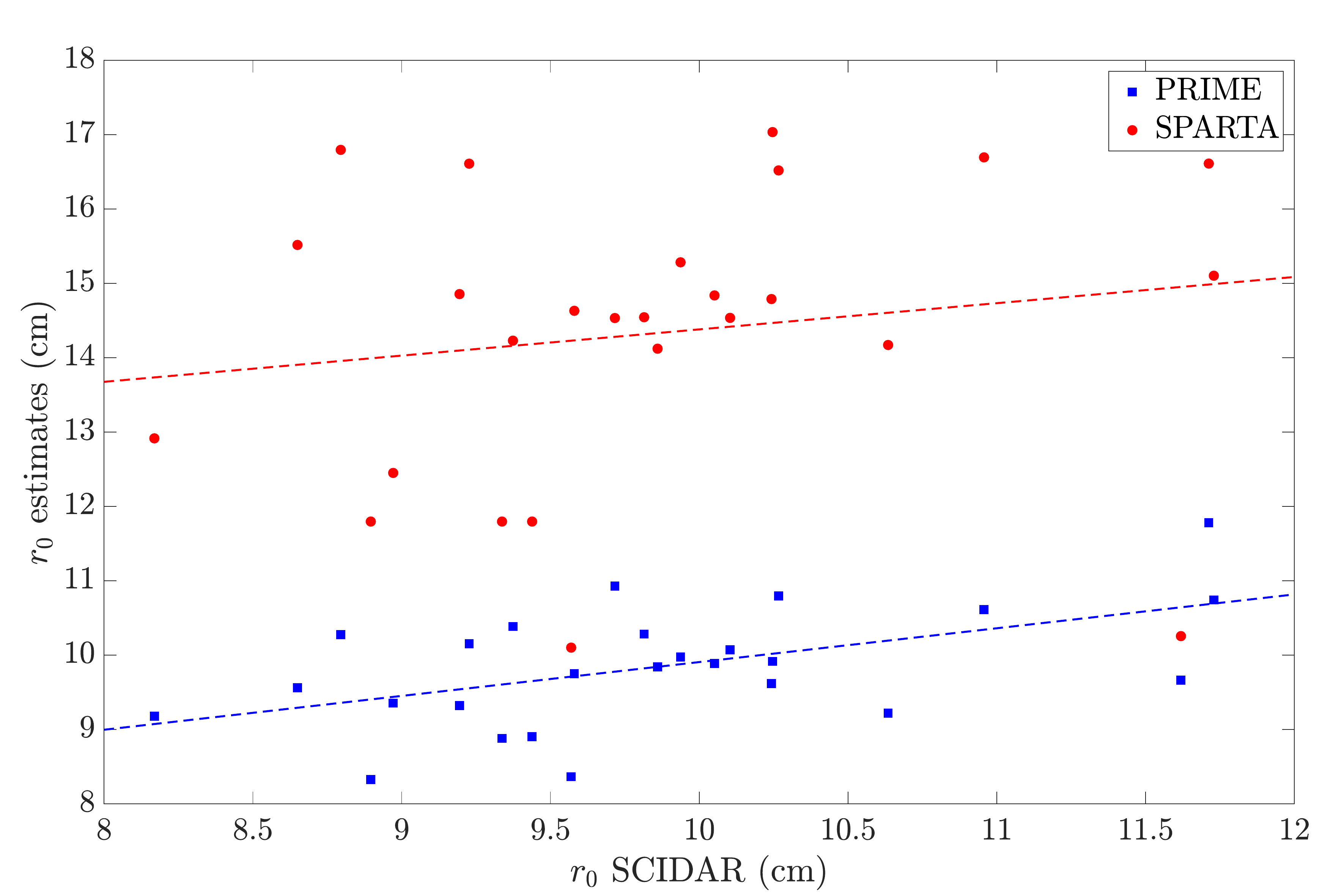}
\caption{\small $\rz$ estimates obtained from SPARTA and PRIME with respect to SCIDAR $\rz$.}
\label{F:r0estVr0scidar}
\end{figure}

\cobm{Furthermore, we have compared $\rz$ estimates provided by both SPARTA (AO telemetry) and PRIME as function of SCIDAR measurements as presented in Fig. \ref{F:r0estVr0scidar}. As already observed in Fig. \ref{F:seeing}, we have a large offset between SPARTA and SCIDAR measurements, which no longer appear when comparing with PRIME results, i.e. PRIME and SCIDAR $\rz$ estimates seem to comply during this observing night. This also advocates that SPARTA overestimates the $\rz$ by a significant amount, which can be unbiased through calibration using PRIME over as many data as possible.}

\begin{figure}
\centering
\includegraphics[width = 8.5cm]{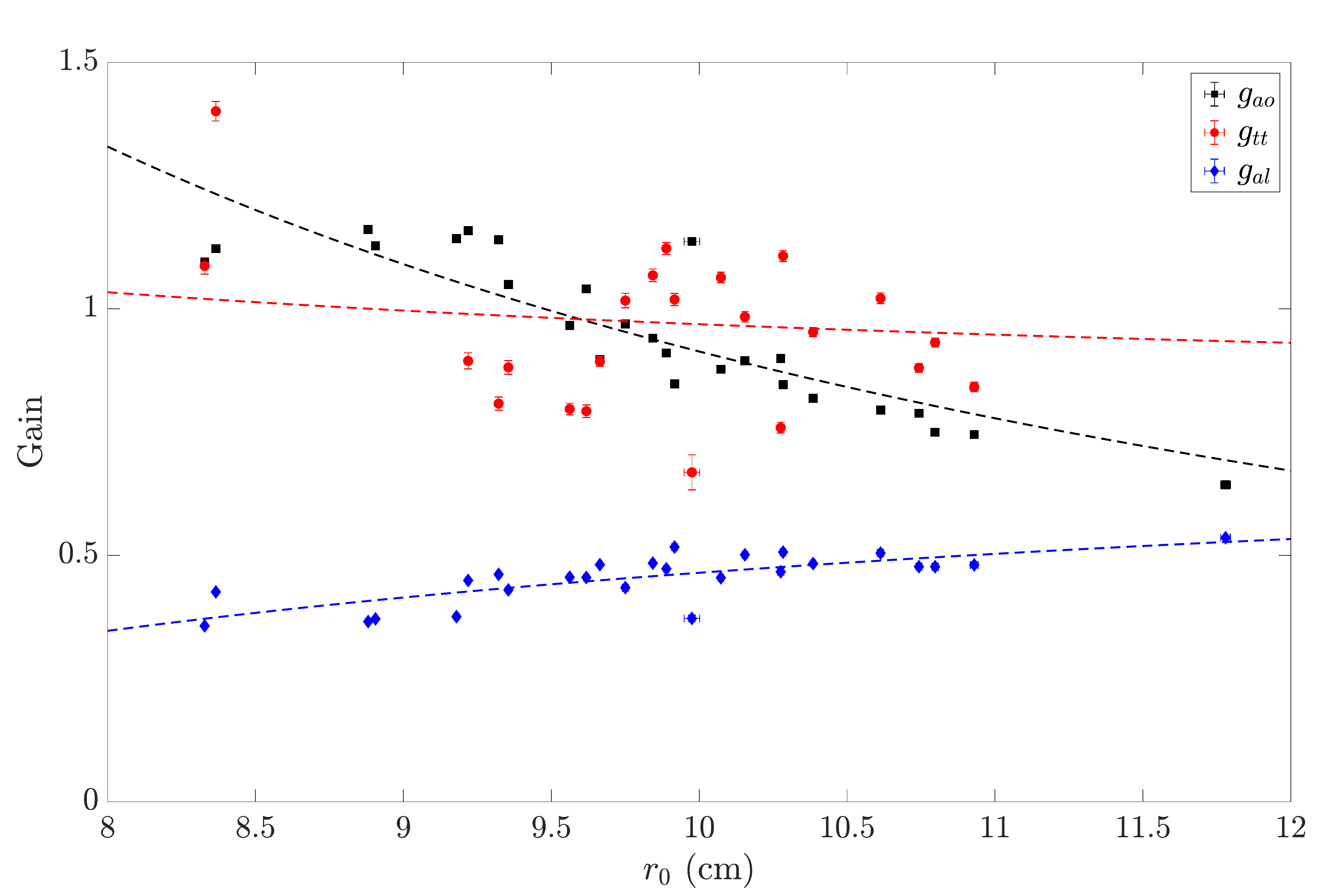}
\caption{\small Retrieved values of $\gao,\gtt$ and $\gal$ as functions of the estimated $\rz$ value with PRIME. Dash line corresponds to trends in $\rz^{-5/3}$ \cobm{using the $\rz$ estimates from with PRIME}. }
\label{F:gainsVr0}
\end{figure}

Finally, we present in Fig. \ref{F:gainsVr0} the evolution of retrieved gains with respect to $\rz$. Identically to what we have done for $\rz$ estimations, gain values are averaged over results obtained with the two AO data sets, and uncertainties show that they were similar and correlated. \cobm{Trends with respect to the $\rz$ estimates from PRIME are obtained using a polynomial fit. These trends allows to visualize how much the parameters vary with respect to the turbulence's strength; they are quite obvious for $\gao$ and $\gal$: when the $\rz$ increases, the turbulence strength and the WFS spot size diminish and consequently the gain value $\gao$ is lessened as well. SPARTA involves a weighted center of gravity algorithm to convert WFS pixels \citep{Petit2008} into WFS slopes, which may introduce optical gain variations across time as well.
Aliasing gain slightly increases with respect to $\rz$, e.g. for weaker turbulence, which also corresponds to shorter turbulence time (so higher wind speed) according to SPARTA measurements in Fig. \ref{F:seeing}. As we previously discussed, the aliasing PSD model is effectively sensitive to turbulence velocity variations though modifications of the aliasing transfer function $\mathcal{H}_\text{cl}$ given in Eq. \ref{E:psdalias}. This one is derived from the AO closed-loop temporal transfer function \citep{Gendron1994} that is converted into a spatial transfer function by turn the temporal frequencies to spatial frequencies using the wind speed value. Higher wind speed values tend to introduce more spatial filtering and therefore decrease the aliasing energy. Consequently, PRIME must boost $\gal$ in order to match the PSF as we observe in Fig. \ref{F:gainsVr0}.
Furthermore, $\gtt$ seems to slightly vary with respect to $\rz$ as well, although the trend is not completely clear. We stress that in V-band the Strehl ratio is about 2\% and the AO performance is dominated by atmospheric residuals, particularly residual tip-tilt. \cred{This latter is influenced by both the seeing and turbulence velocity, especially for those observations were the system was running at 300\,Hz. Fig. \ref{F:PSFs} illustrates that the PSF is elongated in the 45 degree direction, certainly due to the combination of the turbulence velocity and AO system servo-lag. } In other words, $\gtt$ variations must be driven by atmospheric turbulence's temporal properties as well.
More generally, discrepancies around the $\rz^{-5/3}$ trends are trackers to detect that $\covTT$ structure changes across time. As we handle the AO telemetry acquired unsynchronously with the imager frame, we compensate the atmospheric turbulence properties by scaling a multiplicative factor that apply to those covariance matrices, i.e. we assume that theirs structures remain identical, only the amplitude changes. PRIME will always retrieve the parameters set that provides the best match with the observed PSF; in other words, these degrees of freedom will absorb any variability of the covariance matrix structure. If we want to use PRIME as a PSF-fitting facility for PSF determination purpose, there is no absolute need to built the PSF model from synchronous telemetry as PRIME will manage to restore the best PSF model for the observed data set. However, if one cares about the physical meaning of retrieved parameters, we must (i) refine the PSF model as accurate as possible by including all physical effects (static aberrations, cophasing errors, WFS optical gains,...) (ii) confront a PSF model to synchronous observations to mitigate temporal drifts of observing conditions that could be absorbed into the parameters fitting as observed in Fig. \ref{F:gainsVr0}.}



To summarize PSF estimation performance, we report in Tab. \ref{T:SRacc} SR and PSF FWHM estimation accuracy. We compare forward PSF-R and PRIME: we highlight that (i) PRIME strongly unbiases estimation of PSF metrics, although some improvement can be pursued on the FWHM, by including the WFS noise model for instance, (ii) PRIME permits to reduce the relative error in average by a factor 7 and 2 on respectively the SR and FWHM. Note that the standard-deviation is also increased by the quite small amount of data we have, the image noise that contaminates the PSF calibration and the very poor seeing conditions. in this respect, Tab. \ref{T:SRacc} provides best results on a worst-case scenario.

\begin{table}
\centering
\caption{\small Final accuracy on Strehl ratio and FWHM obtained with either forward PSF-R (no model adjustment) or PRIME.}
\begin{tabular}{|c|c|c|c|c|}
\hline
& \multicolumn{2}{|c|}{$\Delta$ SR (\%)} & \multicolumn{2}{|c|}{$\Delta$ FWHM (mas)}  \\
\hline
& bias & std & bias &std\\
\hline
Forward PSF-R slot 1 & 27 & 12 & 4.2 & 2.1 \\
\hline
Forward PSF-R slot 2 & 12 & 21 & 3.5 & 4.0 \\
\hline
PRIME slot 1 & 0.2 & 2.9 & -0.4 & 0.9 \\
\hline
PRIME slot 2 & -0.4 & 2.2 & -0.8 & 1.6 \\
\hline
\end{tabular}
\label{T:SRacc}
\end{table}

\begin{figure}
\centering
\includegraphics[width = 8.5cm]{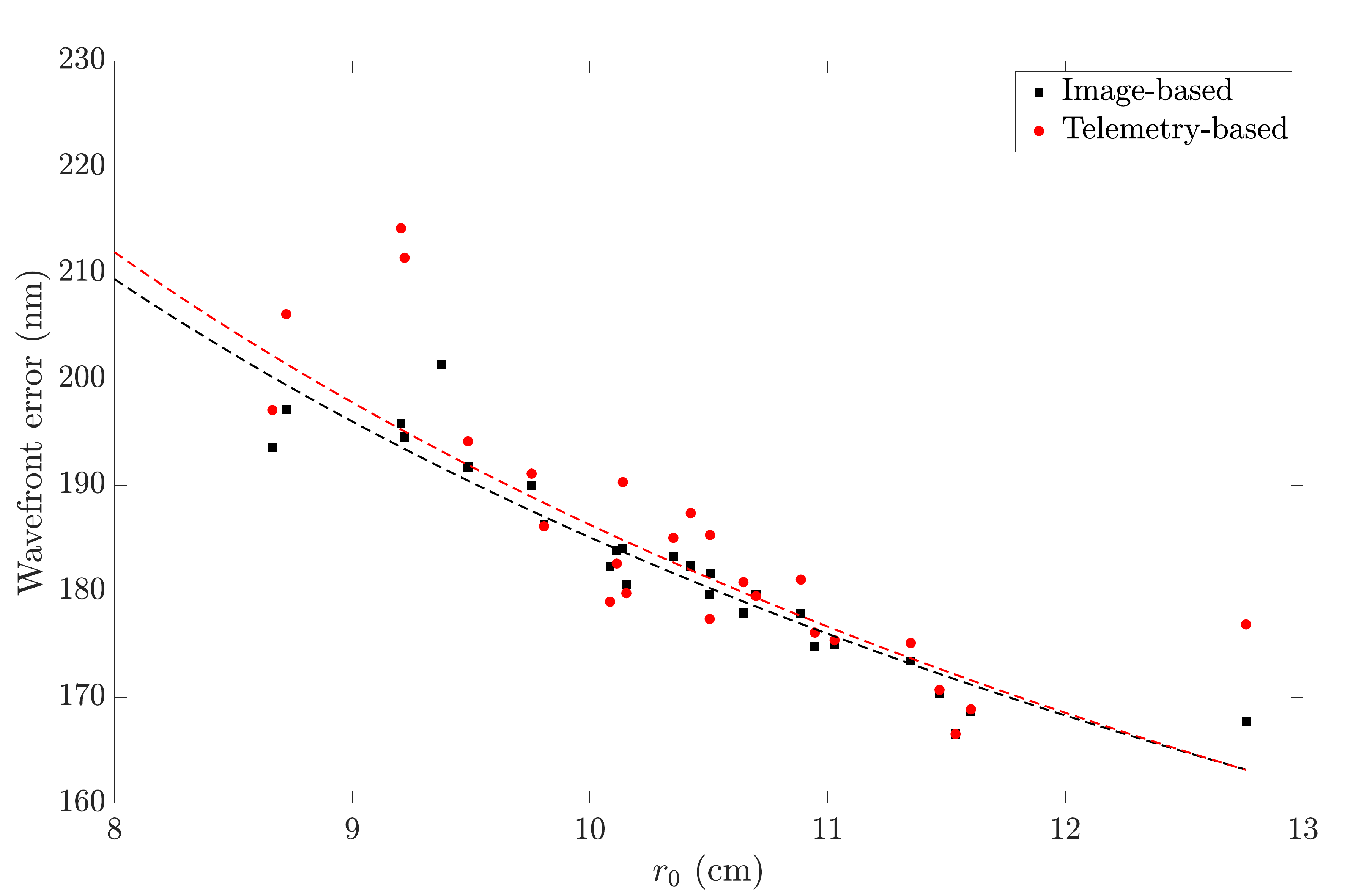}
\caption{\small Residual wavefront error with respect to $\rz$ calculated either from the image SR or the AO telemetry and the PRIME-adjusted parameters. Dash lines give trends in $\rz^{-5/6}$ \cobm{using the $\rz$ estimates from with PRIME}. }
\label{F:wfeVr0}
\end{figure}

In order to provide more confidence into the retrieval process, we have built a \textit{a posteriori} AO error breakdown to obtain a SR values to compare with the image measurements. We estimate the residual error from
\begin{equation}
\sigma^2_\varepsilon = \sigma^2_\text{ao} + \sigma^2_\text{tt} + \sigma^2_\perp + \sigma^2_\text{noise},
\end{equation}
where
\begin{itemize}
\item [$\bullet$] \cobm{$\sigma_\text{ao} = \sqrt{\gao\times \text{tr}(\mathcal{C}_\text{AO})/1377}$} is the tip-tilt excluded AO residual. This is assessed to 135\,nm$\pm$10\,nm over the full observation.
\item [$\bullet$] \cobm{$\sigma_\text{tt} = \sqrt{\gtt\times \text{tr}(\mathcal{C}_\text{TT})}$}. We have retrieved 190\,nm $\pm$ 18\,nm.
\item [$\bullet$] \cobm{$\sigma_\perp = \sqrt{0.2(d/\rz)^{5/3}}$} is the DM fitting error that reached 69\,nm$\pm$5\,nm.
\item [$\bullet$] \cobm{$\sigma_\text{noise}$} is estimated on open loop slopes including tip-tilt and was measured at the level of 46\,nm$\pm$18\,nm.
\end{itemize}
We present in Fig.\ref{F:wfeVr0} the residual wavefront error determined from either the SR image or the error breakdown described above. Thanks to PRIME, AO parameters are estimated accordingly to the final residual error we must retrieve. On top of the PSF fitting, this gives another evidence that parameters found by PRIME are connected to the system performance.

PRIME offers a new opportunity: instead of acquiring synchronous AO control loop data with science observations, one may record AO telemetry time to time, along a sufficient time to make atmosphere characteristics converges, probably at least 30\,s \citep{Martin2012}. However, this demands to investigate few questions, which are (i) how frequent we should acquire AO data, in other words how frequent covariance functions structure changes? (ii) can we find proper descriptors (seeing, turbulence velocity, telescope elevation)  to track structure changes? (iii) can we enable accurate forward PSF-R by inferring what should be $\gao,\gtt,\gal$ and $\rz$ from contextual data (AO telemetry, external $\cnh$ profilers, telescope data)? Provide insights on these questions will be next step of this work.

\section{Application to faint stars}
\label{S:offaxisPRIME}

\cobm{So far, we have deployed PRIME on the on-axis bright star, with very good conditions of S/N and absence of crowding. However, PRIME can provide accurate PSF in more challenging conditions, where other alternatives are not feasible. We aim in this section to test the PSF model calibration using off-axis stars, which are much fainter (mV = 15-16\,mag). One must consequently rely on a PSF-fitting algorithm to recover the PSF morphology and this situation is ideal to test PRIME capabilities in such a scheme. Besides, we have highlighted in Sect. \ref{SS:offaxisPSFR} that the PSF is \cred{shift-invariant} across the field. In other words, the ground-truth is given by the on-axis PSF; the best performance PRIME can achieve on present images are provided in Sect. \ref{SS:onaxisPRIME}. Nevertheless, in order to mitigate the noise propagation in our fitting procedure, we must regularize the criterion presented in Sect. \ref{E:crit}. We have followed the present strategy:
\begin{itemize}
\item[$\bullet$] \textbf{Step 1:}  PSF model parameters $\rz, \gao, \gtt, \gal$ are calibrated using off-axis stars \#1 and \#2 with different regularization strategies presented in Sect. \ref{SS:reg}. We obtain eventually a calibrated 2D PSF model.
\item[$\bullet$] \textbf{Step 2:} The resulting PSF model is used to retrieve the on-axis star's photometry and astrometry, and the final accuracy depends on the PSF fitting performance during step 1. \cred{Eventually, we assess a mean square error using Eq. \ref{E:res}.} Thanks to results presented in Sect. \ref{SS:onaxisPRIME}, we analyze how the different regularization strategies affect the PSF fitting residual and the PSF parameters identification.
\end{itemize}}

\subsection{Regularization}
\label{SS:reg}

We have tested two different regularization strategies:
\begin{itemize}
\item[$\bullet$] \textbf{Bounds regularization:} We minimize the criterion presented in Eq. \ref{E:crit} by limiting the parameter space with strict boundaries, inside which the parameter probability distribution is uniform. Bounds are defined relatively to $\boldsymbol{\mu}_0$, the solution we have retrieved in Sect. \ref{SS:onaxisPRIME}:\cred{
\begin{equation}
\label{E:JB}
 J_B(\boldsymbol\mu,\gamma,\boldsymbol\alpha) = \left\{\begin{array}{ll}
     J(\boldsymbol\mu,\gamma,\boldsymbol\alpha) & \text{if $\Vert\boldsymbol\mu - \boldsymbol\mu_0\Vert \le \boldsymbol\sigma_\mu$} \\
     +\infty & \text{else} \\
   \end{array}\right.  
\end{equation}}
\cred{where the constrain is component-wise as $\boldsymbol{\mu}$ includes several parameters}. Therefore, if bounds are set to 0 (no uncertainty) we retrieve the PSF we obtain in our first scenario in Sect. \ref{SS:onaxisPRIME} where we use the central star to adjust those parameters. By enlarging bounds around the optimal solution, we make the minimization process more sensitive to the image noise and check how the solution deviates from the optimal one.\newline  
\item[$\bullet$] \textbf{Gaussian regularization:} \cred{we assume that PSF parameters follow a Gaussian distribution $\mathcal{N}(\boldsymbol\mu_0, \boldsymbol\sigma_\mu)$}, where $\boldsymbol\mu_0$ is the prior on parameters and $\boldsymbol{\sigma}^2_\mu$ the distribution variance. In order to regularize the problem, we update the criterion given in Eq. \ref{E:crit} by adding a regularization term on the PSF parameters as follows:
\begin{equation}
\label{E:JG}
\mathcal{J}_\text{G}(\boldsymbol{\mu},\gamma,\boldsymbol{\alpha}) = \mathcal{J}(\boldsymbol{\mu},\gamma,\boldsymbol{\alpha}) + \Gamma\sum_{i=1}^{n_\mu} \norme{ \dfrac{\boldsymbol{\mu} - \boldsymbol{\mu}_0}{\boldsymbol{\sigma}_\mu}}^2,
\end{equation}
where $n_\mu$ is the number of adjusted PSF parameters (we have $n_\mu=4$ in the present analysis) and $\Gamma$ a scalar factor that weights the contribution of the regularization term. The present formulation is convenient to separate the parameters precision $\boldsymbol{\sigma}_\mu$ that is identified from calibration (optical gains) of an estimation process ($\rz$) from $\Gamma$, which is a user-defined factor that reflects how much we are confident into the $\boldsymbol{\mu}_0$ a priori by balancing the weight between the data-based and regularization terms. For instance, if one apply PRIME over successive frames, $\boldsymbol{\sigma}_\mu$ will be delivered for each processed frame as a confidence interval on retrieved parameters that depends on the sole frame only. Thus, we obtain a new $\boldsymbol{\mu}_0$ vector for the next frame, and depending how much correlated are those two frames, the user may decide to give more weight to the regularization (boost $\Gamma$) or not. Potentially, we can have very precise model-fitting, but the found parameters won't necessarily represent the PSF when treating another frame acquired few minutes later due to observing conditions variations. On the contrary, we could also have badly precise estimates due to poor S/N conditions, but stationarity of the observing conditions. Having both parameters $\boldsymbol{\sigma}_\mu$ and $\Gamma$ allows to discriminate PSF parameters precision and accuracy.\newline

However, one must understand that regardless the regularization, we are going to bias the PSF fitting results by forcing the solution space to stick around $\boldsymbol{\mu}_0$. The bounds regularization means that the parameters Probability Density Function (pdf) is a gate function and is relevant when we have physical constrains to limit the solution space. Thus, we expect this strategy to propagate more noise, i.e. the PSF fitting residual will degrade, with respect to the pdf width. On top of that, the pdf is uniform over the space of acceptable solutions, which means that we do not specifically trust or give more weight to the initial set that feeds PRIME for instantiating the first iteration. The Gaussian regularization differs one this point: we must have some confidence in the prior meaningfulness to choose such a regularization.\newline
Our purpose is now to identify how PRIME behaves with respect to (i) the regularization strategy, either uniform pdf over a bounded space or Gaussian (ii) the degrees of freedom of each of those, i.e. $\boldsymbol{\sigma}_\mu$ (both of them) and $\Gamma$ (Gaussian pdf only). 
\end{itemize}

\subsection{PSF fitting results}
Firstly, we have tried PRIME without any regularization which led to an inefficient reconstruction with a mean residual error (Eq. \ref{E:res}) of 3\%, while we obtained about 0.1\% in Sect. \ref{SS:onaxisPRIME}. The $\rz$ and $\gal$ parameters estimates were particularly biased and physically meaningless, which calls for regularization.
\cred{In order to provide concise an readable results, we do not have set the parameters precision independently to each others. Instead, we have made vary $\boldsymbol{\sigma}_\mu$ relatively, from 0\% to 90\%, to $\boldsymbol{\mu}_0$.  Each parameter pdf width will be absolutely different, but will have the same relative width regarding the optimal solution. Eventually, this methodology will permit to define which overall parameters precision is required to achieve a given accuracy on the PSF.} Besides, for the Gaussian regularization, we have also tested different values of $\Gamma$ ($0.001$ up to $0.1$). \cred{The prior $\boldsymbol\mu_0$ in Eq. \ref{E:JG} has been set to the optimal solution found from PRIME on the on-axis image in Sect. \ref{SS:onaxisPRIME}. Therefore, if we choose a too large value for $\Gamma$, the criterion in Eq. \ref{E:JG} will be dominated by the regularization term and the solution will stick to $\boldsymbol\mu_0$, i.e. we must retrieve the residual error we have got in Sect. \ref{SS:onaxisPRIME}. On the contrary, if $\Gamma$ is too low, the regularization term has not impact on the solution and we will propagate too much noise in the solution and obtain meaningless results. The point of this analysis is then to identify which range of $\Gamma$ must be envisioned to improve results compared to the bounds regularization.}

\begin{figure}
\centering
\includegraphics[width = 8.5cm]{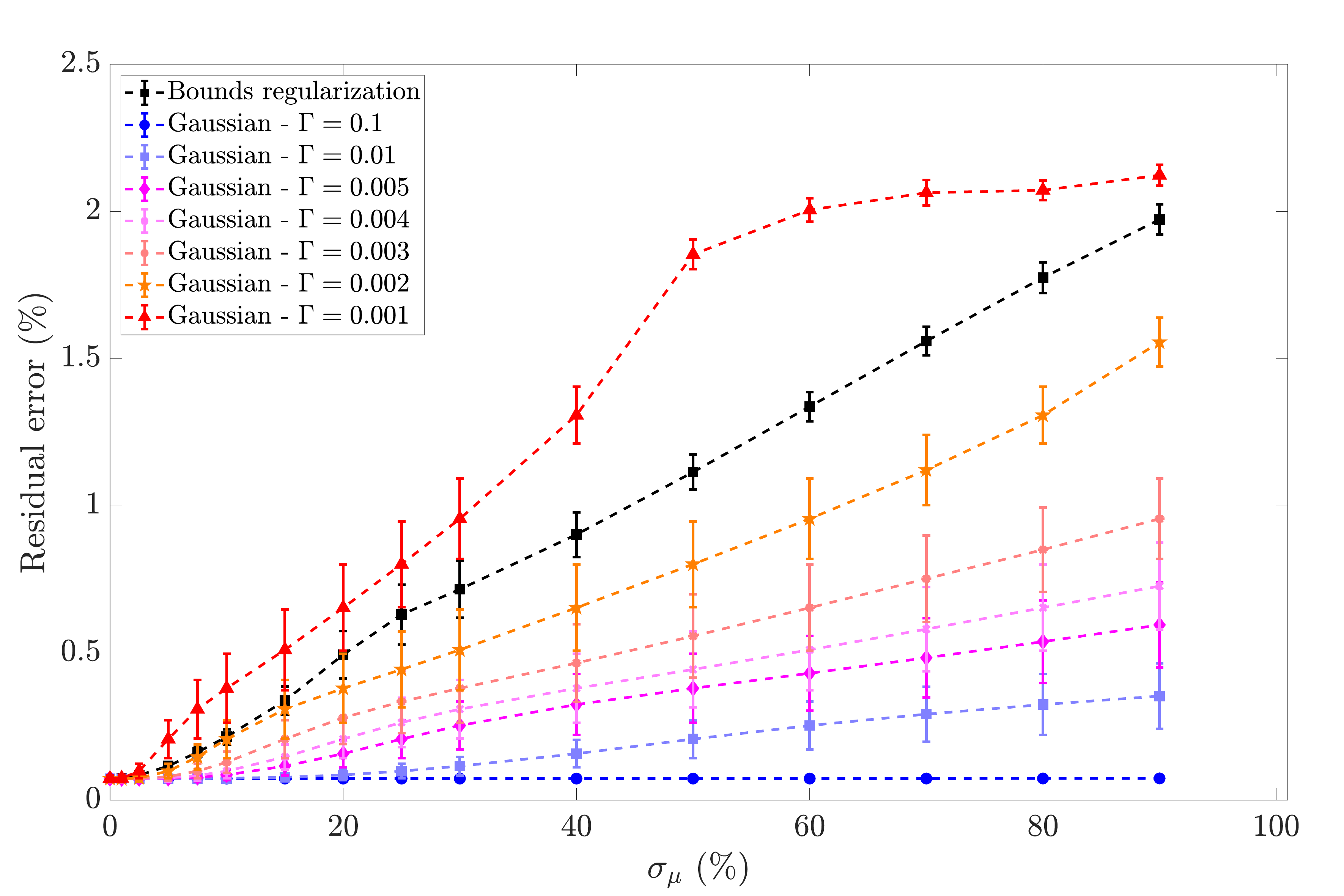}
\caption{\small Mean square error as calculated in Eq. \ref{E:res} as function of $\boldsymbol{\sigma}_\mu$ for several regularization strategies. Errors bars are given by averaging values over the 12 first ZIMPOL frames. }
\label{F:eqmVinterval}
\end{figure}

We present the residual error in Fig. \ref{F:eqmVinterval}. \cred{Residual errors are calculated by comparing the reconstructed PSF (step 2) with the on-axis image using Eq. \ref{E:res}.} As expected, wider pdf ($\boldsymbol{\sigma}_\mu$ larger) conducts to more noise propagation and worse PSF fitting results. Using the bounds regularization, the residual error grows up quite linearly with respect to $\boldsymbol{\sigma}_\mu$ and we may be able to maintain the residual error below 1\% by containing parameters within 40\% from the ground truth, while the Gaussian regularization allows to control the slopes of this degradation. According to Fig. \ref{F:gainsVr0}, PSF parameters may varies significantly across time, but variations seem to remain within 50\% for this night. If AO telemetry is systematically available with synchronous focal-plane images, these variations should be controlled as the telemetry signal scales directly with the real $\rz$ values. In other words, the precision we must consider is given by the standard-deviation of parameters estimates from the $\rz^{-5/3}$ trends presented in Fig. \ref{F:gainsVr0}, which stands up to 10-20\%, meaning that PRIME is able to maintain the mean residual error below 0.5\% if synchronous telemetry is provided.

\begin{figure}
\centering
\includegraphics[width = 8.5cm]{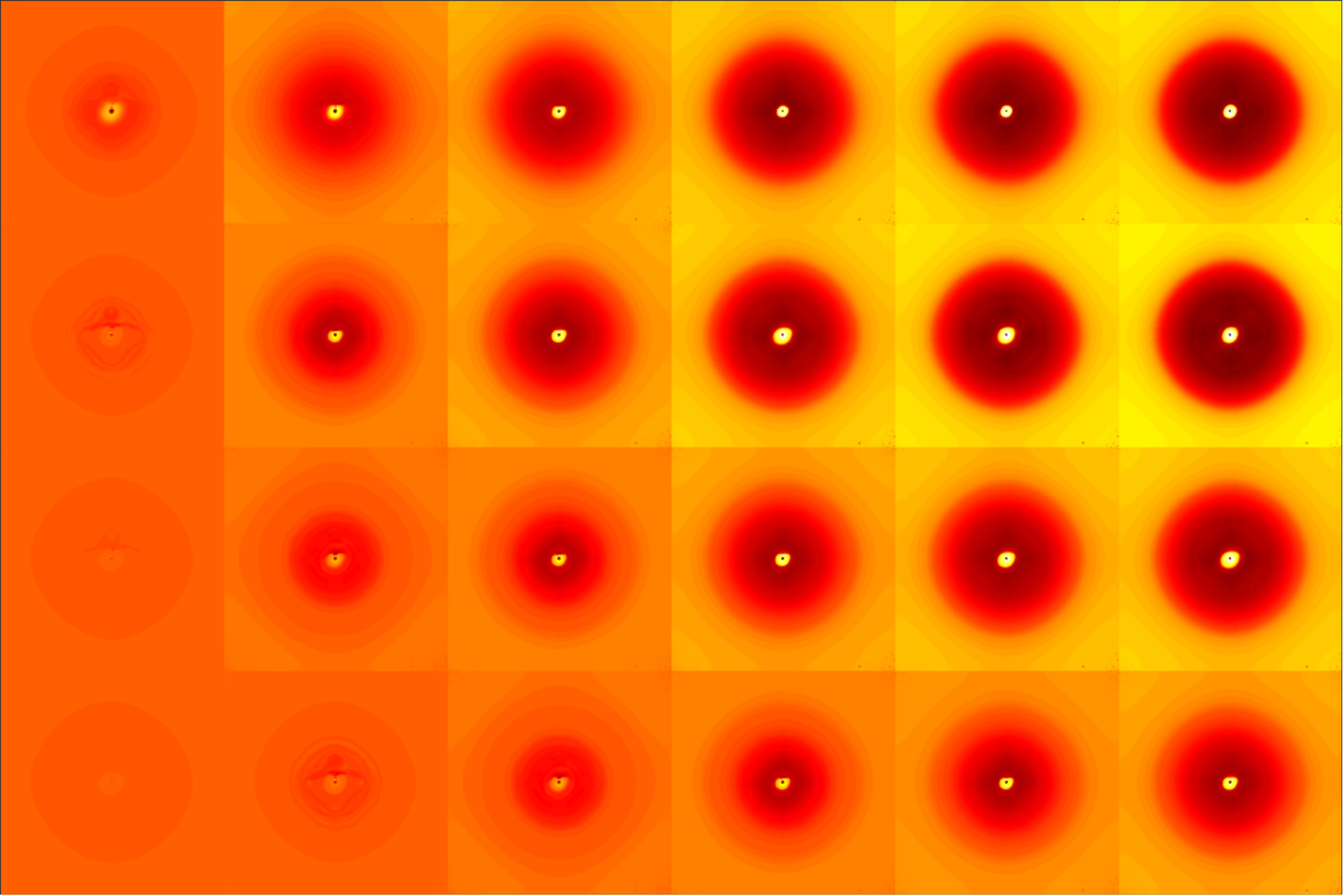}
\caption{\small On-axis PSF fitting residual obtained after calibration using off-axis stars and displayed in hyperbolic arcsinus (model underestimation in black) scale. \textbf{From left to right} $\boldsymbol{\sigma}_\mu =$ 1, 5 10, 20, 30 and 40\,\%. \textbf{From top to bottom:} bounds regularization, Gaussian regularization with $\Gamma = $ 0.002, 0.005 and 0.01.}
\label{F:psfsVbounds}
\end{figure}

\cobm{Fig. \ref{F:psfsVbounds} illustrates how the PSF fitting degrades with respect to $\boldsymbol{\sigma}_\mu$ and $\Gamma$. We notice that the inner AO corrected region is firstly overestimated for low values of $\boldsymbol{\sigma}_\mu$. For larger values, the residual is mostly dominated by the non-corrected spatial frequencies with an underestimation of the PSF wings (overestimation of the $\rz$) and an overestimation of the background in Eq. \ref{E:crit}. In other words, adjusted parameters degrade in opposite direction due to noise confusion, in a way the total energy (or flux) is relatively well conserved while the PSF structure is not.}

\begin{figure}
\centering
\includegraphics[width = 8.5cm]{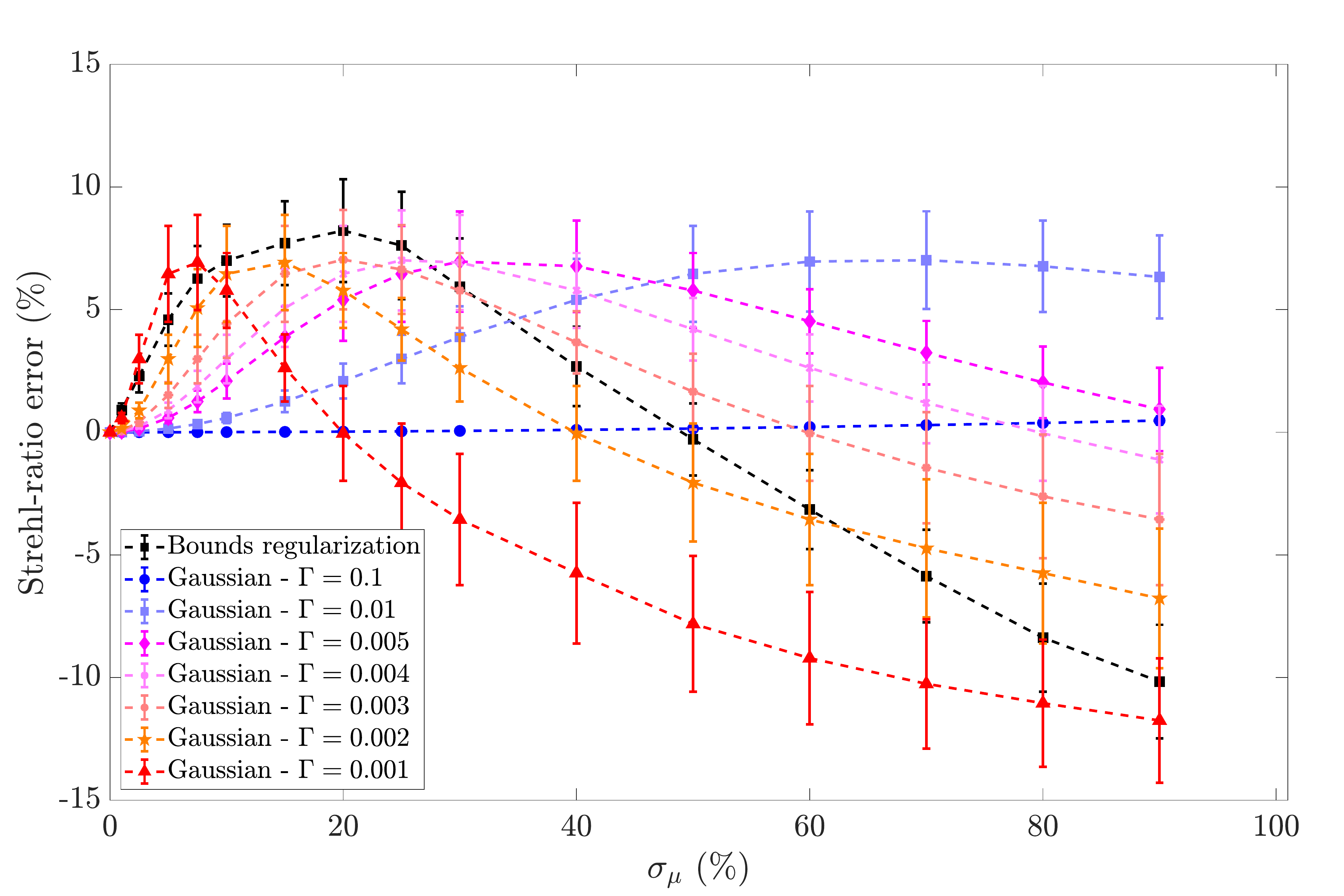}
\caption{\cobm{\small Strehl-ratio error with respect to $\boldsymbol{\sigma}_\mu$ using different regularization strategies.}}
\label{F:dSRVbounds}
\end{figure}

\begin{figure}
\centering
\includegraphics[width = 8.5cm]{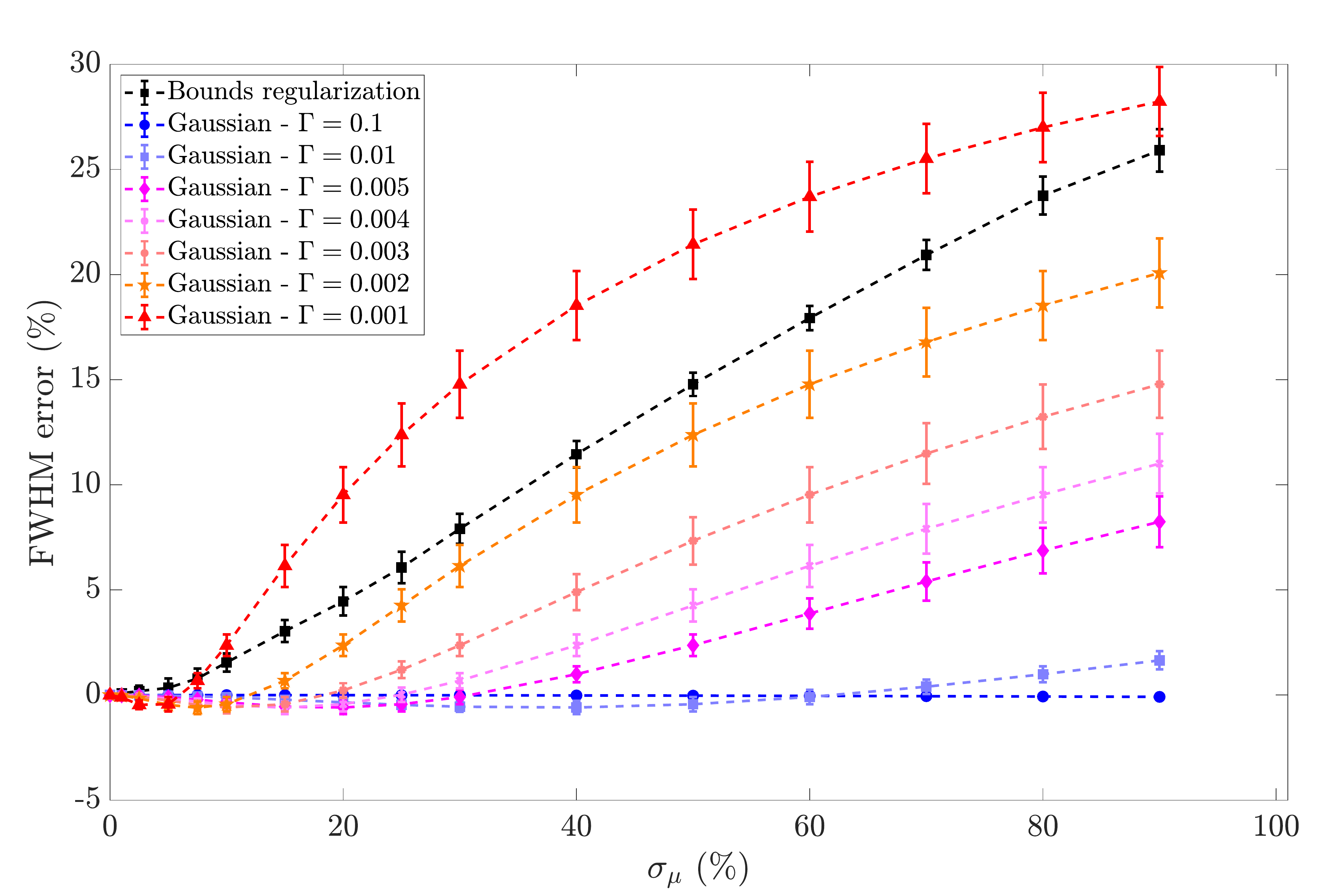}
\caption{\cobm{\small FWHM error with respect to $\boldsymbol{\sigma}_\mu$ using different regularization strategies.}}
\label{F:dFWHMVbounds}
\end{figure}

\cobm{Fig. \ref{F:dSRVbounds} and \ref{F:dFWHMVbounds} present the SR and FWHM estimation error as function of $\boldsymbol{\sigma}_\mu$ and $\Gamma$. We retrieve similar behavior compared to what we have shown in Fig. \ref{F:eqmVinterval}. The FWHM estimates is getting worse, but remains within 10\% as long as the initial guess remains within 40\% from the truth. The SR error evolves differently; first it increases and finally drops down to negative value: there is a switch that occurs at a $\boldsymbol{\sigma}_\mu$ value that depends on $\Gamma$. Below this threshold, the noise propagation through the criterion solving shifts the solution from he ground truth but without modifying significantly the PSF morphology (the FWHM is reconstructed at few percents). Above this threshold value, PRIME seems to fall into a different local minimum by injecting too much energy in the background and not enough in the PSF wings. However, SR error stays below 10\% despite the PSF structure is badly reconstructed, which advocates for exploring alternatives metrics for PSF fitting quality assessment.}

\begin{figure}
\centering
\includegraphics[width = 8.5cm]{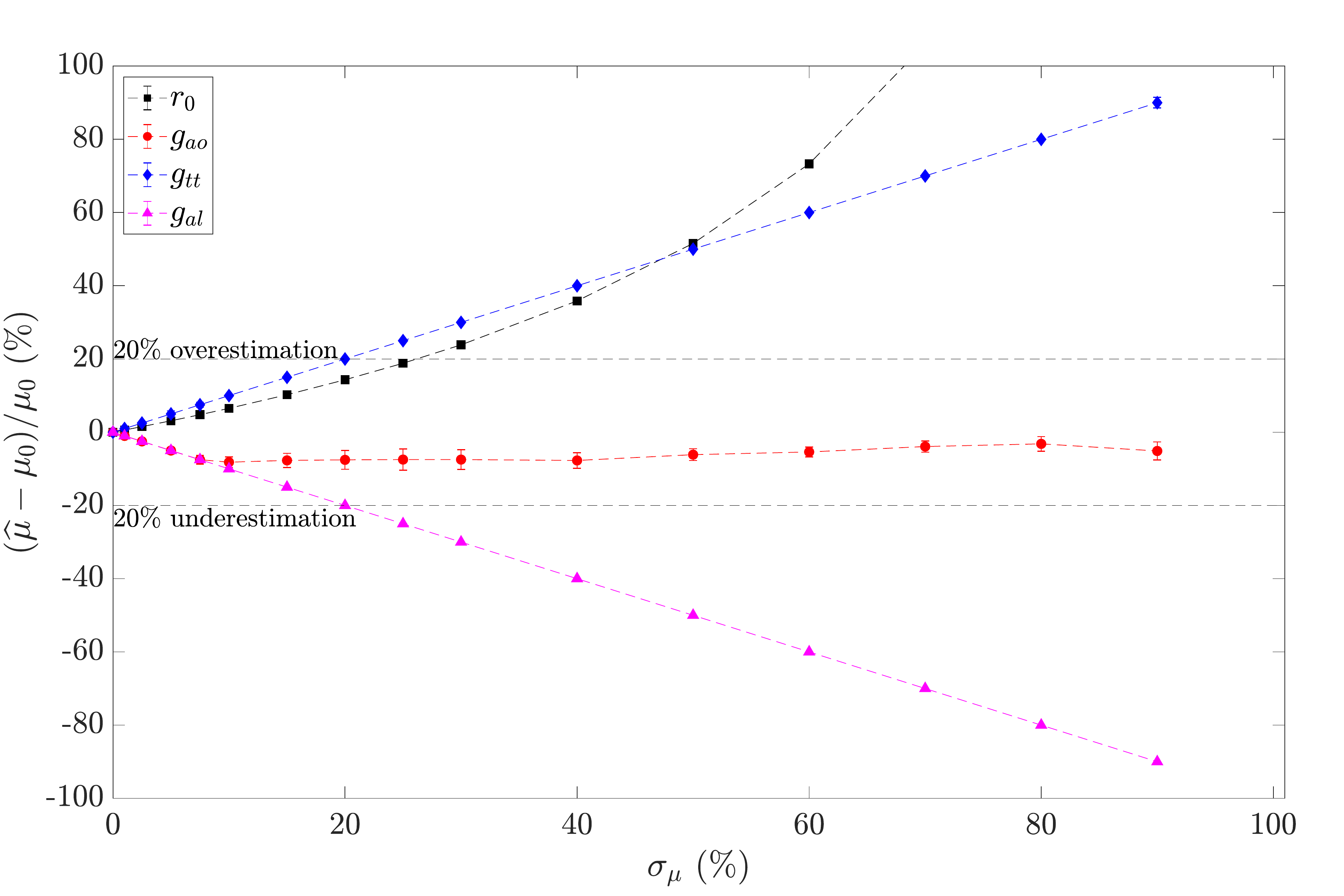}
\includegraphics[width = 8.5cm]{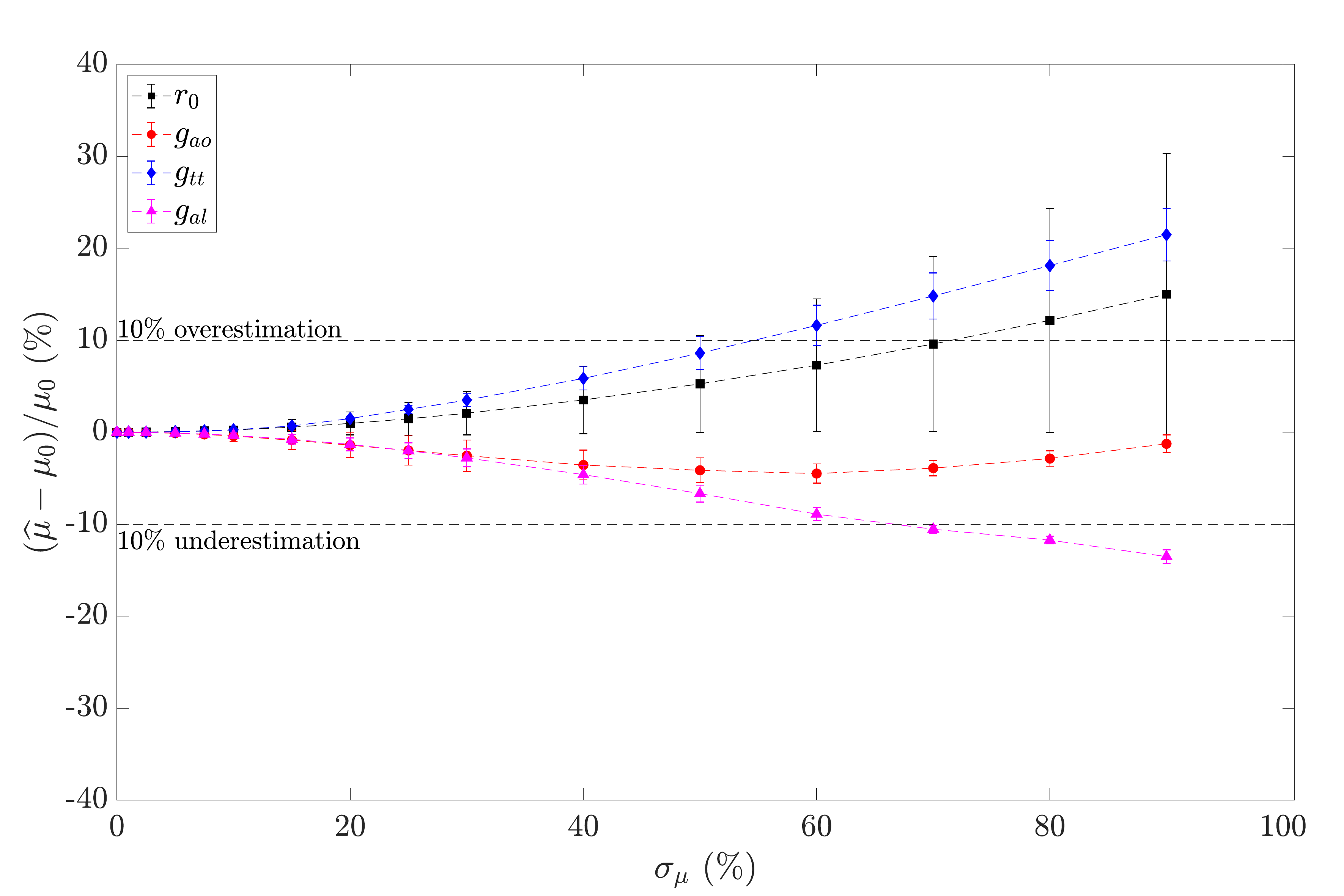}
\caption{\cobm{\small  Estimation error on PSF model parameters obtained with PRIME on off-axis stars with respect to $\boldsymbol{\sigma}_\mu$ using different regularization strategies: (\textbf{Top:}) Bounds regularization (\textbf{Bottom}) Gaussian regularization with $\Gamma=0.005$.}}
\label{F:errVbounds}
\end{figure}

\cobm{Previous conclusions are confirmed when looking in Fig. \ref{F:errVbounds} that shows the retrieved parameters as function of $\boldsymbol{\sigma}_\mu$ and regarding several regularization strategies. The bounds regularization shows that $\gal$ and $\gtt$ degrades linearly with respect to $\boldsymbol{\sigma}_\mu$. The noise confusion brings already a PSF degeneracy on parameters retrieval, but those ones are quite orthogonal, i.e. their impact on the PSF is significantly different. However, the $\rz$ varies quadratically, which reinforces the idea that PRIME confuses the energy between the background and the PSF wings that speeds up the $\rz$ estimates degradation. On the contrary, $\gao$ remains well estimated within 10\% as it refers to the AO corrected area, where the S/N is maximal. This is also reassuring to confirm that the $\gao$ estimation is decoupled from others PSF parameters.}

\cobm{ Regarding PSF fitting and parameters estimation results, we distinguish three different regimes:
\begin{itemize}
\item[$\bullet$] $\boldsymbol{\Gamma <0.002}$: the problem is under-regularized and the fitting residual error increases with respect to $\boldsymbol{\sigma}_\mu$ rapidly up to the worse case scenario (no regularization). This configuration must be avoided in favor of the bounds regularization if $\boldsymbol{\mu}_0$ is not trusted or if we only know physical constrains to bound the solution space.
\item[$\bullet$] $\boldsymbol{0.002\leq \Gamma < 0.01}$: there is a good balance between data-based and regularization terms and this configuration should be preferred to the bounds regularization, as long as we have some confidence into the priors $\boldsymbol{\mu}_0$, i.e. we have some evidence that the solution must be close to $\boldsymbol{\mu}_0$. Such confidence can be brought by processing successive frames, or using models of spatial PSF variations to perform the fitting on good S/N PSF first before treating fainter stars scenarios.
\item[$\bullet$] $\boldsymbol{0.01\leq \Gamma}$: the problem is over-regularized and the adjusted parameters remain very close to $\boldsymbol{\mu}_0$, i.e. the solution is highly biased. This configuration should only serve to slightly update a solution that has been determined over an image highly correlated with the current processed.
\end{itemize}}
\cred{Regarding these results, we may envision to perform the PSF fitting with PRIME in a sequentially manner:  (i) use bright PSFs in the field to constrain the model with a bounds regularization and obtain a preliminary set of parameters with associated precision, (ii) redo this process starting from the retrieved parameters as an initial guess and a prior $\boldsymbol\mu_0$ with a Gaussian regularization and $\Gamma = 0.002$ (iii) repeat step 2 with larger and larger $\Gamma$ value until reaching the best precision on parameters. On top of that, we can apply step 2 on fainter stars to refine the model sequentially using more information without propagating too much noise. }

\section{Conclusion}
\label{S:conclusion}

In this article, we presented an innovative application of the PSF-R method to SPHERE/ZIMPOL data. This is a crucial experiment since the diffraction limit achieved in optical with a 10-meter class of telescope is comparable to ELTs in near infrared. Thus, this kind of studies paves the road for exploiting ELTs data reduction.

We have first described how we perform classical PSF-R using SPHERE control loop data. Unfortunately, the current PSF-R framework do not allow us to achieve a proper and stable PSF estimation. In order to be improved, this would require to push deeply our AO system understanding through several calibration, thus demanding a substantial amount of observing and/or technical time.

To overcome this problem, we have introduced the PRIME approach, that is a PSF-fitting technique that inherits from the PSF-R framework to calibrate the PSF model. This is initially instantiated from the AO control loop data. We have shown that PRIME allow us to achieve a very accurate PSF modeling at better than 0.1\% of mean residual, with AO telemetry unsynchronized from scientific images; acquiring 30\,s of AO data every hour was sufficient for obtaining excellent results with PRIME.

\cobm{Finally, we have tested PRIME over faint stars to compare the calibrated PSF to the ground-truth given by the on-axis image. This was possible since, as verified using stereo-SCIDAR data, the anisoplanatism does not contribute significantly to the PSF morphology. In order to enable a meaningful PSF estimation in such severe S/N conditions, we have presented two different strategies to regularize the minimization criterion, either using a truncated but uniform pdf or a Gaussian one. PRIME ensures to obtain 1\% of mean residual error on the PSF by bounding the solution space within 40\% from the optimal solution. The Gaussian regularization allows to increase estimates accuracy up to the optimal achievable performance depending on how much we trust PSF parameters prior. The Gaussian regularization allows the user to adapt the data-based/regularization terms balance in the minimization criterion regarding how much priors on parameters are trusted. Furthermore, thanks to the PSFR framework, the spatial PSF variations can be accurately modeled, which offers alternative possibilities for PSF-fitting problems (1) use all external information we have to define PSF parameters priors (Stereo-scidar for instance) (2) instantiate the PSF model calibration on available PSFs with good S/N and poor crowding using a bounds regularization (3) use the retrieved information plus variations models to fit the PSF on fainter stars using a Gaussian regularization (4) repeat step 2 and 3 with updated information on the PSF using a Gaussian regularization until reaching the minimal residual. }

We plan to push this work further by collecting more science observations, including crowded stellar field. When a large sample of data sets will be at disposal, we will apply statistical inference tools in order to capture how PSF model parameters vary with respect to contextual data. This will enable forward PSF estimation in case of lack of point sources in the field and enhance PRIME efficiency for stellar fields applications.

The next step will be to plug PRIME within a standard image analysis pipeline in order to combine the strengths of both PSF-R techniques and multi-source image processing tools. This is a necessary step in order to allow a large diffusion of the use of the PSF-R which is nowadays only confined to very few experts.

\section*{Acknowledgments}
Authors acknowledge the French National Research Agency (ANR) to support this work through the ANR APPLY (grant ANR-19-CE31-0011) coordinated by B. Neichel. G.F. has been supported by the Futuro inRicerca 2013 (grant RBFR13J716).


\bibliographystyle{mnras} 
\bibliography{biblioLolo}
\bsp
\label{lastpage}	
\end{document}